\documentclass{article}

\usepackage{graphicx} 
\usepackage{url}
\usepackage{tikz}
\usepackage{tikz-qtree,tikz-qtree-compat}
\usepackage{subcaption}
\usepackage{amssymb}
\usepackage{amsmath}
\usepackage{float}
\usepackage{array,colortbl}

\definecolor{mplblue}{RGB}{31,119,180}
\definecolor{mplorange}{RGB}{255,127,14}

\newcolumntype{O}{>{\columncolor{mplorange}}c}
\newcolumntype{B}{>{\columncolor{mplblue}}c}

\providecommand{\keywords}[1]
{
  \small	
  \textbf{\textit{Keywords---}} #1
}

\title{Stringology-Based Motif Discovery from EEG Signals: an ADHD Case Study}

\author{Anat Dahan, Samah Ghazawi\thanks{Corresponding author: samahi@braude.ac.il}\\
 \small Braude College of Engineering, Karmiel, Israel\\}
\date{}
\begin{document}
\maketitle

\begin{abstract}
We propose a novel computational framework for analyzing electroencephalography (EEG) time series using methods from stringology, i.e., the study of algorithmic techniques for efficiently processing and analyzing strings, which enables the systematic identification and characterization of recurrent temporal patterns in neural signals. The goal of this work is to introduce new measures for understanding neural signal dynamics, with the reported findings serving as a proof-of-concept demonstration of the framework. The framework adapts order preserving matching (OPM) and Cartesian tree matching (CTM) to discover temporal motifs that preserve relative ordering and hierarchical relationships while remaining invariant to amplitude scaling, providing a temporally precise representation of EEG dynamics complementary to spectral and global complexity analyses. 

To demonstrate its utility, we applied the framework to multichannel EEG recordings from individuals with attention-deficit/hyperactivity disorder (ADHD) and matched controls, a publicly available dataset. Highly recurrent, group-unique motifs (support $\ge 0.9$) were identified and quantified using OPM and CTM. ADHD participants showed significantly higher motif frequencies, indicating increased repetitiveness in neural activity. OPM analysis revealed shorter motif lengths and greater gradient instability for the ADHD group, reflected in larger mean and maximal inter-sample amplitude jumps. CTM analysis corroborated these findings and further revealed reduced hierarchical complexity in ADHD, characterized by shallower tree structures and fewer hierarchical levels, despite comparable motif lengths.

These results indicate that ADHD-related EEG alterations manifest as systematic differences in the structure, stability, and hierarchical organization of recurrent temporal patterns, extending beyond spectral or global complexity measures. The proposed stringology-inspired motif framework provides a complementary computational tool for EEG analysis, with potential applications in objective biomarker development and monitoring of neurodevelopmental disorders.
\end{abstract}


\section{Introduction}
\label{intro}
Attention-deficit/hyperactivity disorder (ADHD) is characterized by atypical neural dynamics that manifest as increased variability, reduced temporal stability, and disrupted coordination across brain networks. Traditional electroencephalography (EEG) analyses often summarize activity over long time windows, potentially overlooking short-lived but functionally meaningful temporal patterns. Capturing these fine-grained recurrent micro-patterns can provide a more precise view of the neural dysregulation underlying ADHD.

Electroencephalography (EEG) provides a direct, high-temporal-resolution measure of neural activity, capturing both ongoing oscillations and transient event-related responses~\cite{cohen2017does}. In ADHD, alterations in the raw EEG signal have been consistently reported across both resting-state and task-related conditions. Early work highlighted increased theta (4-8Hz) and decreased beta (13-30Hz) activity, yielding the classic "theta/beta ratio" biomarker, often interpreted as reflecting cortical underarousal and attentional dysregulation~\cite{ahn2025adhd,barry2009eeg,hasler2016attention}. More recent studies have demonstrated that these spectral abnormalities are spatially and developmentally variable, suggesting that ADHD is better conceptualized as involving atypical temporal organization and dynamic reconfiguration of neural networks rather than a single frequency imbalance~\cite{lenartowicz2014use,loo2012clinical,loo2018parsing,luo2023aberrant}.

Beyond oscillatory power, analyses of event-related potentials (ERPs), such as the P3, N2, and contingent negative variation (CNV), have revealed attenuated or delayed responses to cognitive and inhibitory cues in ADHD, reflecting deficits in executive control, response monitoring, and attentional allocation~\cite{doehnert2010mapping}. These transient, phase-locked EEG features complement spectral findings by highlighting disruptions in stimulus-driven processing speed and amplitude modulation.

However, traditional ERP and spectral analyses reduce the rich temporal complexity of the EEG to either averaged waveforms or power estimates, potentially overlooking subtle yet meaningful temporal structures. To address this limitation, several approaches have been developed to analyze the raw EEG signal itself, that is, its evolving temporal morphology rather than summary statistics. Nonlinear methods such as Lempel-Ziv complexity (LZC), approximate entropy (ApEn), and fractal dimension analysis have been applied to capture the irregularity and unpredictability of EEG time series in ADHD, consistently reporting reduced complexity indicative of less integrative and more stereotyped brain activity~\cite{ccetin2022case,lau2022brain,zarafshan2016electroencephalogram}. Other frameworks, such as microstate analysis, decompose the EEG into transient spatial patterns that reflect quasi-stable network configurations, revealing shorter microstate durations and altered switching dynamics in ADHD~\cite{berchio2025eeg}.

Despite these advances, existing analytic approaches often summarize neural activity using global averages, whether spectral power, complexity indices, or ERP amplitudes, without characterizing the specific temporal patterns underlying these measures. As a result, the precise sequence and structural organization of neural fluctuations remains poorly understood. Traditional methods quantify the variability or irregularity of the EEG signal but do not identify the types of temporal motifs that occur, their frequency of repetition, or structural differences between populations. 

To address this gap, the present study introduces methods adapted from stringology, i.e., the study of algorithmic techniques for efficiently processing and analyzing strings. Specifically, order preserving matching (OPM)~\cite{opm} and Cartesian tree matching (CTM)~\cite{ctm1} frameworks are used to analyze EEG dynamics in ADHD. These pattern matching approaches preserve the relative ordering and hierarchical structure of EEG time-series segments, respectively, enabling comparisons of temporal patterns independent of amplitude scaling. By applying motif discovery~\cite{doi:10.1142/9781860947292_0011}, i.e., the identification of highly recurrent temporal patterns, we discover recurring structures in the EEG signal based on OPM and CTM. Comparing these motif characteristics between the ADHD and control groups allows us to examine differences in temporal organization, stability, and structural complexity that may reflect dysregulation in attentional and executive control networks. 
Specifically, this study seeks to:
\begin{itemize}
\item[(1)] Identify motif-based features that distinguish EEG dynamics in ADHD from those in typical EEG.
\item[(2)] Examine how these motifs relate to known measures of temporal stability and complexity.
\item[(3)] Establish motif-based analysis as a complementary framework to conventional spectral and nonlinear methods for understanding the temporal architecture of ADHD brain activity.
\end{itemize}
Importantly, the primary goal of this work is to introduce new measures for understanding neural signal dynamics, with the reported findings serving as a proof-of-concept demonstration of the framework’s analytical potential rather than as definitive clinical conclusions.

The paper is organized as follows: first, we describe the dataset collected during an experimental task in Section~\ref{data}. Second, we define the analytic tools, i.e., OPM and CTM, in Section~\ref{string}. This is followed by a detailed implementation of preprocessing and analysis steps in Section~\ref{methods}. Next, we present the results in Section~\ref{results}. The final sections, i.e., Sections~\ref {discussion} and~\ref {conclude}, focus on discussion and conclusions, respectively.

\section{Dataset}
\label{data}
The dataset\footnote{The dataset is available at https://dx.doi.org/10.21227/rzfh-zn36} consists of EEG recordings from 121 children (boys and girls) aged 7-12 years, divided into two groups, ADHD and control~\cite{nasrabadieeg}. The ADHD group includes 61 children diagnosed according to the DSM-IV criteria~\cite{widiger1997dsm}, all of whom had been under Ritalin treatment for up to six months before the EEG recording. The control group consists of 60 typically developing children with no reported history of psychiatric disorders, epilepsy, or high-risk behaviors.

EEG signals were acquired using a 19-channel system (see Table~\ref{tab:eeg_channels}) arranged according to the standard 10-20 electrode placement scheme, with reference electrodes (A1 and A2) positioned on the earlobes. The recordings were sampled at 128Hz, and the duration of each EEG session varied according to the child’s response time during the experimental tasks. Before analysis, the EEG signals were preprocessed to reduce noise and remove artifacts.

\begin{table}[H]
\centering
\small
\begin{tabular}{|l|l|l|}
\hline
\textbf{Brain Region} & \textbf{Electrodes} & \textbf{Channels} \\
\hline
Frontal   & Fp1, Fp2, F3, F4, F7, F8, Fz & 1, 2, 3, 4, 11, 12, 17 \\
Central   & C3, C4, Cz & 5, 6, 18 \\
Temporal  & T7, T8 & 13, 14 \\
Parietal  & P3, P4, P7, P8, Pz & 7, 8, 15, 16, 19 \\
Occipital & O1, O2 & 9, 10 \\
Reference & A1, A2&---\\
\hline
\end{tabular}
\caption{EEG electrode groups and corresponding channel numbers. This configuration provided comprehensive coverage of frontal, central, temporal, parietal, and occipital brain regions.}
\label{tab:eeg_channels}
\end{table}

Participants completed a visual attention task. Images of cartoon characters were presented, and the children were instructed to count the number of characters in each image.
The number of characters in the images ranged randomly from 5 to 16, ensuring that the task was engaging and visually accessible to the participants.
EEG data collection was performed during the task, with each image displayed following the participant's response, creating a dynamic and interactive recording session.

The dataset is designed to study brain connectivity, attention mechanisms, and the nonlinear dynamics of EEG signals. It provides a valuable resource for investigating diagnostic methods and computational approaches for ADHD.
\section{Stringology Tools}
\label{string}
A string is a finite sequence of characters defined over a well-specified alphabet. For example, if the alphabet is $\{a,b,c\}$, then the sequence $cbbacabcc$ is a valid string over this alphabet. Research on algorithms and combinatorics applied to strings is known as \textit{stringology}, a term coined by Zvi Galil in 1985~\cite{galil}. 

Stringology provides algorithmic methods for analyzing, manipulating, and discovering patterns in strings. It enables the detection of hidden regularities, simplification of data representations, and the efficient analysis of sequential information. Applications of stringology include information retrieval, computational biology, bioinformatics, natural language processing, cryptography, and signal processing~\cite{book}.  

In this study, we employ three stringology-based tools: motif discovery, order preserving matching (OPM), and Cartesian tree matching (CTM).

\subsection{Motif Discovery}
Motifs are recurring or statistically significant substrings within a collection of strings~\cite{doi:10.1142/9781860947292_0011}. They capture structural or functional regularities and constitute a fundamental concept in stringology. 

For example, consider the set of strings
\[
S = \{abad, xabaxx, cqxxzababaw, dxxcaba\}.
\]
The substring $aba$ occurs in all strings in $S$, and thus forms a motif of the group with a support value of $100\%$. While the substrings $a$, $b$, $ab$, and $ba$ also occur in all strings, we consider only the maximal substring $aba$, as it subsumes these shorter occurrences.
Additionally, the substring $xx$ occurs in four strings, i.e., in the majority of the group. Consequently, $xx$ can be considered a motif of $S$ with a support value of $75\%$. In general, motif discovery aims to identify recurring substrings along with their frequencies and supports within the dataset.

\subsection{Order Preserving Matching (OPM)}
\label{opm}
Order preserving matching (OPM) is a pattern matching paradigm that compares strings based on the relative ordering of their elements while ignoring absolute values~\cite{opm}. OPM is particularly suited for numerical strings, where relative trends are more informative than precise magnitudes.
Formally, two strings $p$ and $t$ of equal length $n$ are said to form an \emph{order preserving match} if, for all $i,j \in \{1,\ldots,n\}$,
\[
p[i] \leq p[j] \iff t[i] \leq t[j].
\]
That is, the pairwise ordering relations between elements in $p$ and $t$ must be identical.
For example, the strings $(2,9,4)$ and $(1,3,2)$ form an OPM match because both follow the same ordinal pattern: first element $<$ third element $<$ second element.

In OPM-based motif discovery, motifs are defined as substrings that recur across a collection of strings while preserving exact ordinal relationships between their elements. Unlike traditional motif discovery, which relies on exact symbol matching, OPM motifs capture recurring relative trends such as consistent increases, decreases, or local extrema.
Due to its strict definition, OPM-based motif discovery identifies only highly stable and repeatable ordinal patterns. Any deviation in the relative ordering of values prevents a match, making OPM motifs particularly precise but less tolerant to noise or small fluctuations.

\subsection{Cartesian Tree Matching (CTM)}
\label{ctm}
Cartesian tree matching (CTM) is a shape-based pattern matching approach that represents a numerical string as a Cartesian tree, capturing its hierarchical structure of local minima and maxima~\cite{ctm1}. This representation allows CTM to identify similarities between strings even in the presence of minor ordinal variations.

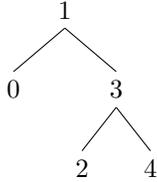
\begin{figure}[H] 
\centering 
\begin{tikzpicture}[level 1/.style={level distance=1.3cm},level 1/.style={sibling distance=5mm},level 2/.style={sibling distance=5mm}] \Tree [.1 [.0 ] [.3 [.2 ] [.4 ] ] ] 
\end{tikzpicture} 
\caption{The Cartesian trees of $(7,2,16,4,8)$ and $(10,5,17,9,18)$ are identical, illustrating a Cartesian Tree Matching (CTM) instance. In both cases, the tree nodes are labeled by the positions of the corresponding values in the string, starting from position $0$. More generally, any string of length $5$ in which the minimum value occurs at position $1$ and the second minimum occurs at position $3$ will produce the same Cartesian tree structure, regardless of the order relations between the values at positions $0$, $2$, and $4$.} 
\label{fig:ctmmatch} 
\end{figure}

A Cartesian tree $CT(p)$ for a string $p$ is defined recursively~\cite{CT}:
\begin{itemize}
    \item If $p$ is empty, then $CT(p)$ is empty.
    \item Otherwise, let $p[i]$ be the leftmost minimum element in $p$. The root of $CT(p)$ corresponds to $p[i]$, its left subtree is $CT(p[1 \ldots i-1])$, and its right subtree is $CT(p[i+1 \ldots n])$.
\end{itemize}
In other words, the smallest value in the string becomes the root of the tree, and the same construction rule is applied recursively to the substrings on either side.
Two strings $p$ and $t$ are said to form a CTM match if their Cartesian trees have identical shapes, regardless of the actual node labels. For example, the strings $(7,2,16,4,8)$ and $(10,5,17,9,18)$ share the same Cartesian tree structure, as illustrated in Figure~\ref{fig:ctmmatch}.

CTM-based motif discovery identifies motifs as recurring substrings whose Cartesian tree representations share the same structural shape. Unlike OPM, CTM allows small variations in local ordering as long as the hierarchical arrangement of minima and maxima is preserved. As a result, CTM motifs capture broader waveform structures rather than exact ordinal patterns.

The relationship between OPM and CTM is hierarchical. Every OPM match induces an identical Cartesian tree and therefore constitutes a CTM match. However, the converse does not necessarily hold: two strings may share the same Cartesian tree structure while differing in their exact ordinal relationships. Consequently, OPM identifies a strict subset of the patterns captured by CTM.

In EEG analysis, stringology-based methods enable the identification of recurrent temporal micro-patterns that reflect underlying neural dynamics. By transforming numerical EEG signals into strings and searching for repeated substrings, it becomes possible to characterize temporal organization, stability, and structural regularities in neural activity. Such properties are particularly relevant in the study of ADHD, which has been associated with increased neural variability and disrupted temporal coordination.

Motif discovery in EEG requires similarity measures that are robust to inter-subject variability and amplitude differences. Methods based on relative ordering or structural shape, rather than absolute values, are therefore well-suited for this domain. In this study, OPM is used to detect highly consistent temporal motifs that preserve exact ordinal relationships between consecutive samples. These motifs correspond to strictly recurring rising and falling trends in EEG signals.
In addition, CTM is employed to identify motifs that preserve the overall waveform shape while allowing minor ordinal variations. This flexibility enables the detection of broader recurrent neural patterns that may vary slightly in local ordering but retain their essential structural characteristics.

Together, OPM and CTM provide complementary perspectives on EEG dynamics. Their combined use enables a nuanced comparison between the ADHD and control groups by capturing both highly stable ordinal motifs and more flexible structural waveform patterns.
The following section describes how OPM- and CTM-based motifs are systematically extracted and quantified from EEG recordings to identify group-specific neural signatures.

\section{Implementation}
\label{methods}
Our analysis proceeded through a systematic pipeline designed to discover and characterize meaningful patterns, i.e., motifs, from the EEG data. We began by organizing the dataset hierarchically, examining each experimental group's recordings independently. Within each channel, we processed the individual sequences through a standardized preprocessing workflow that included high-pass and low-pass filtering to remove artifacts and noise, Independent Component Analysis (ICA) to separate signal sources, and finally, 1-in-4 downsampling was performed to reduce computational complexity while preserving essential frequency information.

Once the signals were preprocessed, we considered each sequence as a string (per group, per participant, and per channel), i.e., each preprocessed EEG channel time series from an individual participant was treated as a separate string. In total, we have $61\times19+60\times19 = 2299$ strings to consider in the motif discovery phase. For each string, we systematically explored maximal motifs, defined as motifs that are not strictly contained within any longer motif, of varying lengths $>5$ (motifs of length $<5$ were excluded, as they were insufficient to capture meaningful temporal structure).
At each length scale, we applied two complementary mining approaches as mentioned previously: OPM motifs and CTM motifs. To ensure we captured only robust and frequently occurring motifs, we set a minimum support threshold of $0.9$ for both mining methods, meaning that discovered motifs had to appear in at least $90\%$ of the analyzed strings. To ensure group specificity, we retained only motifs that did not appear in the other group. This stringent criterion ensures we identify only robustly recurrent patterns that represent characteristic features of group-level neural dynamics rather than idiosyncratic patterns present in small subsets of individuals. Lower thresholds would increase false discovery of spurious patterns due to individual variability, while higher thresholds for support would overly restrict the analysis to near-universal patterns.

\subsection{Motif Feature Definitions}
We extracted four categories of features from discovered motifs: initial characteristics, gradient dynamics, rank-based ordinal properties, and hierarchical tree structures.
\paragraph*{Motif Initial Features}
For both types, i.e., OPM motifs and CTM motifs, the following initial features were obtained:
    \begin{itemize}
        \item \textbf{Length}: Length of the motif string. 
        \item \textbf{Frequency}: How many times this motif occurs in all strings within the group. 
        \item \textbf{Support}: The number of strings containing the motif divided by the total number of strings within the group.
        \item \textbf{Mean}: The position-wise average amplitude computed across all instances of motifs.
    \end{itemize}

To characterize the temporal dynamics and structural properties of identified OPM and CTM motifs, we consider a comprehensive set of gradient characteristic features. These features capture different aspects of motif behavior, based on their fine-grained temporal evolution, enabling robust discrimination between EEG patterns across groups. Let \[x=(x_1,\ldots,x_n),\] represent the motif, i.e. $x_i$ is the amplitude value at position $i$ of the motif $x$.

\paragraph*{Motif Gradient Features}
These features capture the dynamics of signal transitions, reflecting the rate and variability of change within motifs:
Let \[J_i = |x_{i+1}-x_i|,\] for $1\leq i\leq n-1$, i.e. the difference between adjacent values in the motif $x$.
\begin{itemize}
    \item \textbf{Max jump}: Maximum absolute difference between consecutive points: \[\max\{J_i|1\leq i\leq n-1\},\]. This captures the most dramatic transition in the motif. For example, in $(10, 12, 20, 18)$, the max jump is $|20-12| = 8$.

    \item \textbf{Mean jump}: Average gradient magnitude, \[\bar{J} = \frac{1}{n-1}\sum_{i=1}^{n-1}J_i,\] representing typical transition steepness.

    \item \textbf{Std jump}: Standard deviation of jump magnitudes, defined formally as \[\sigma^J = \sqrt{\frac{1}{n-1}\sum_{i=1}^{n-1}(J_i - \bar{J})^2},\] measuring variability in the rate of change (smoothness vs. irregularity).

    \item \textbf{Normalized jumps}: Jumps scaled by the motif's amplitude range to enable cross-motif comparison, defined formally as \[J_i^{\text{norm}} = \frac{J_i}{\max(x) - \min(x)},\] for $1\leq i\leq n-1$, where \[\max(x)=\max\{x_i|1\leq i\leq n\},\] and \[\min(x)=\min\{x_i|1\leq i\leq n\}.\] This makes features comparable across motifs with different scales.
\end{itemize}

The same gradient-based features were computed for both OPM and CTM motifs; in the CTM case, these features were extracted directly from the motif before Cartesian tree construction.
Next, we analyze the OPM and CTM motifs, considering each motif type independently and applying different analytical procedures.

\paragraph*{OPM Motif Rank Features}
These features assess the ordinal structure of motifs, making them robust to amplitude scaling and focusing on relative temporal ordering:
Let $r_i$ define the rank of $x_i$ within the motif.
\begin{itemize}
    \item \textbf{Rank diff mean}: Mean absolute difference in ranks between consecutive points, \[\bar{R} = \frac{1}{n-1}\sum_{i=1}^{n-1}|r_{i+1} - r_i|.\] This measures monotonicity, where perfectly monotonic motifs have minimal rank differences. For example, $(5, 2, 8, 1)$ has ranks $(3, 2, 4, 1)$ with rank differences $|2-3|=1$, $|4-2|=2$, $|1-4|=3$, yielding mean = 2.0.

    \item \textbf{Rank diff max}: Maximum rank jump, \[\max\{|r_{i+1} - r_i|\text{ for }1\leq i\leq n-1\},\] capturing the most abrupt change in relative ordering.

    \item \textbf{Rank correlation}: Spearman correlation coefficient between motif ranks and their temporal positions, \[\rho_s = 1 - \frac{6\sum (r_i-i)^2}{n(n^2-1)}.\] High positive correlation indicates a strong monotonic increase; high negative correlation indicates a monotonic decrease.

    \item \textbf{Rank turning points}: Number of indices $2 \le i \le n-1$ at which the direction of rank change reverses, i.e.,
\[(r_i - r_{i-1})(r_{i+1} - r_i) < 0.\]
This feature quantifies oscillatory structure in the motif by counting local reversals in the rank trend. Monotonic motifs have zero turning points, unimodal motifs have a single turning point, while motifs with repeated rises and falls exhibit higher values.
\end{itemize}

Together, the OPM-derived features enabled a comprehensive characterization of EEG temporal dynamics across multiple scales. \textbf{gradient features} quantify local temporal dynamics and signal volatility; and \textbf{rank-based features} provide scale-invariant measures of ordinal structure. This multi-faceted approach allowed us to characterize not only what patterns appear in the EEG signals, but how they evolve, how consistently they occur, and how their structure differs between experimental groups, hence providing a rich feature space for subsequent classification and group discrimination.

\paragraph*{CTM Motif Tree Structural Features}
A Cartesian tree structure maps the hierarchical relationships among local minima and maxima within a segment, where the root of the tree represents the global minimum of the segment, and subsequent branches recursively represent sub-segments. This representation captures the waveform's intrinsic hierarchical organization and shape independent of amplitude scaling, making it particularly robust for comparing patterns across different signal amplitudes or recording conditions.
The following features characterize the topological properties of the Cartesian tree representation of the motif, denoted by $CT$:

\begin{itemize}
    \item \textbf{Tree max depth}: Maximum nesting depth of $CT$, defined as \[d_{\max} = \max(\text{depth}(v)),\] for all nodes $v \in CT$. This measures structural complexity, where deeper trees indicate more hierarchical levels of extrema. For example, a simple oscillation like $(1, 3, 2)$ has depth 2, while a complex nested pattern has greater depth.

    \item \textbf{Tree leaves}: Number of leaves in $CT$, denoted by $N_{\text{leaves}}$. This reflects structural simplicity (few leaves) versus fragmentation (many leaves). A monotonic sequence has a single leaf, while highly oscillatory signals have many.

    \item \textbf{Tree branches}: Number of internal nodes, i.e., non-leaf nodes, denoted by $N_{\text{branches}}$. This counts the subdivisions within the hierarchical structure, indicating the degree of recursive decomposition.

    \item \textbf{Tree balance}: Normalized measure of left/right subtree balance, quantifying symmetry. Computed as \[B = 1 - \frac{|N_L - N_R|}{N_L + N_R},\] where $N_L$ and $N_R$ are the numbers of nodes in left and right subtrees of the root, respectively. Perfect balance yields $B = 1$.
\end{itemize}
For example, in Figure~\ref{fig:ctmmatch}, both trees have $5$ nodes in total, and are of max depth $3$. Additionally, both have $3$ leaves and $2$ internal nodes, i.e., branches. Moreover, $B=1-\frac{|1 - 3|}{1 + 3}=0.5$.

Together, the CTM-derived features provide a comprehensive characterization of EEG waveform structure that is invariant to amplitude scaling. \textbf{Tree structural features} capture the hierarchical organization of extrema, revealing the complexity and symmetry of oscillatory patterns. While \textbf{gradient features} quantify the dynamics of temporal change. This combination enables robust comparison of waveform morphology across different recording conditions and subject groups.

\subsection{Motif Features Statistical Analysis}
To evaluate group-specific differences in EEG motif features between ADHD and control groups, we conducted a comprehensive comparison across all non-initial features and EEG channels. This analysis was performed separately for both OPM-derived and CTM-derived feature sets.

For each group, median feature values were computed to represent central tendencies while mitigating the impact of outliers, which are common in EEG data due to artifacts and biological variability. Group differences were assessed using the Mann-Whitney U test (also known as the Wilcoxon rank-sum test), a non-parametric approach that makes no assumptions about the underlying distribution of the data. This test is particularly appropriate for EEG features, which often exhibit skewed or non-Gaussian distributions.

To control for multiple comparisons across the large number of features and channels tested, we applied False Discovery Rate (FDR) correction to all $p$-values using the Benjamini-Hochberg procedure with a significance threshold of $q < 0.05$. This procedure controls the expected proportion of false positives among all rejected hypotheses, providing a balance between statistical power and Type I error control. Only features that remained significant after FDR correction were considered statistically reliable and included in the final interpretation. The detailed results are presented in the following section.
\section{Results}
\label{results}
The results are presented in the same feature order as defined in the previous section, and separately, for OPM and CTM motifs.
We analyzed motifs with support $\ge 0.9$ for both OPM and CTM approaches, ensuring that only highly recurrent patterns were considered. 
The discovered OPM motifs are of length 6 and 7 (only one motif of length 8 in the ADHD group in channel 1 was discovered), while the discovered CTM motifs are of length 9 and 10, i.e., longer motifs compared to OPM, see Tables~\ref{tab:opmlens} and~\ref{tab:ctmlens} that present tabular visualizations of motif counts per channel and length for OPM and CTM, respectively. Note that the motif counts of each length are mostly larger in the ADHD group compared to the control group.
Motifs are discovered from the downsampled EEG sequences. Since the original signals were sampled at 128Hz and subsequently downsampled using a 1-in-4 scheme, the effective sampling rate of the analyzed sequences is 32Hz. Consequently, each motif element corresponds to $1/32=31.25~ms$ of signal duration. A motif of length $L$ therefore spans $L/32$ seconds; for example, motifs of lengths 6 and 10 correspond to $187.5~ms=0.1875~s$ and $312.5~ms=0.3125~s$, respectively.

\begin{table}[!t]
\centering
\caption{OPM motif counts per channel, per length, and per group. (Orange - ADHD, Blue - Control)}
\small
\begin{tabular}{|c|O|B|O|B|}
\hline
Channel & Len 6& Len 6  &Len 7 & Len 7 \\
\hline
1  & 101 & 3  & 70 & 22 \\
2  & 46  & 8  & 62 & 15 \\
3  & 50  & 6  & 63 & 20 \\
4  & 52  & 10 & 51 & 21 \\
5  & 53  & 3  & 41 & 32 \\
6  & 35  & 6  & 59 & 21 \\
7  & 47  & 4  & 60 & 9  \\
8  & 35  & 6  & 55 & 18 \\
9  & 54  & 5  & 57 & 25 \\
10 & 33  & 16 & 67 & 23 \\
11 & 44  & 14 & 65 & 23 \\
12 & 74  & 6  & 40 & 43 \\
13 & 113 & 0  & 63 & 40 \\
14 & 107 & 8  & 40 & 46 \\
15 & 51  & 7  & 41 & 41 \\
16 & 81  & 3  & 42 & 54 \\
17 & 67  & 6  & 40 & 38 \\
18 & 42  & 3  & 53 & 25 \\
19 & 44  & 9  & 48 & 27 \\
\hline
\end{tabular}
\label{tab:opmlens}
\end{table}

\begin{table}[!t]
\centering
\caption{CTM motif counts per channel, per length, and per group. (Orange - ADHD, Blue - Control)}
\small
\begin{tabular}{|c|O|B|O|B|}
\hline
Channel & Len 9 & Len 9  & Len 10  & Len 10  \\
\hline
1  & 17 & 4 & 85 & 9  \\
2  & 10 & 2 & 62 & 19 \\
3  & 13 & 2 & 56 & 13 \\
4  & 9  & 3 & 76 & 15 \\
5  & 15 & 1 & 71 & 27 \\
6  & 8  & 6 & 57 & 16 \\
7  & 10 & 3 & 75 & 11 \\
8  & 7  & 7 & 57 & 21 \\
9  & 10 & 1 & 68 & 14 \\
10 & 9  & 8 & 61 & 24 \\
11 & 5  & 2 & 64 & 15 \\
12 & 15 & 4 & 66 & 19 \\
13 & 17 & 1 & 82 & 5  \\
14 & 13 & 4 & 50 & 21 \\
15 & 12 & 2 & 59 & 15 \\
16 & 19 & 4 & 58 & 26 \\
17 & 10 & 3 & 71 & 19 \\
18 & 9  & 4 & 67 & 10 \\
19 & 13 & 2 & 57 & 12 \\
\hline
\end{tabular}
\label{tab:ctmlens}
\end{table}

\begin{figure}[H]
    \centering
    \includegraphics[width=1.0\linewidth]{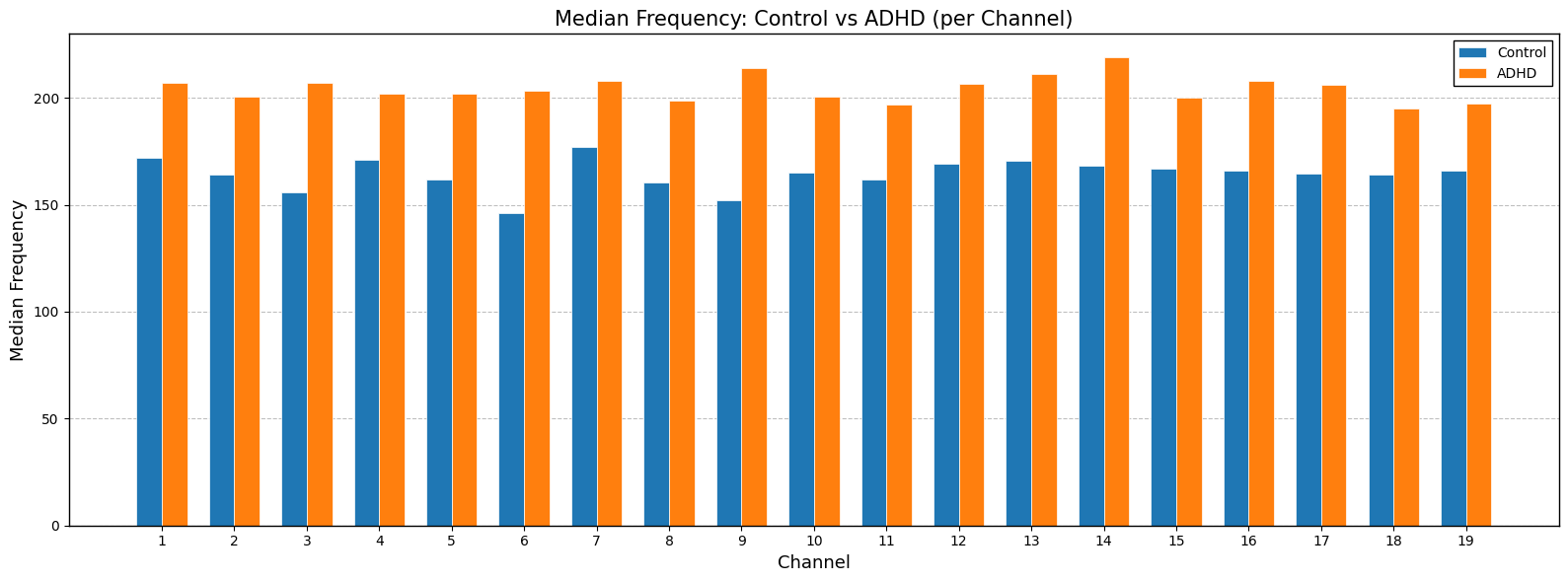}
    \caption{Median frequency of discovered OPM motifs across all EEG channels.}
    \label{fig:opfrequency}
\end{figure}

\begin{figure}[H]
    \centering
    \includegraphics[width=1.0\linewidth]{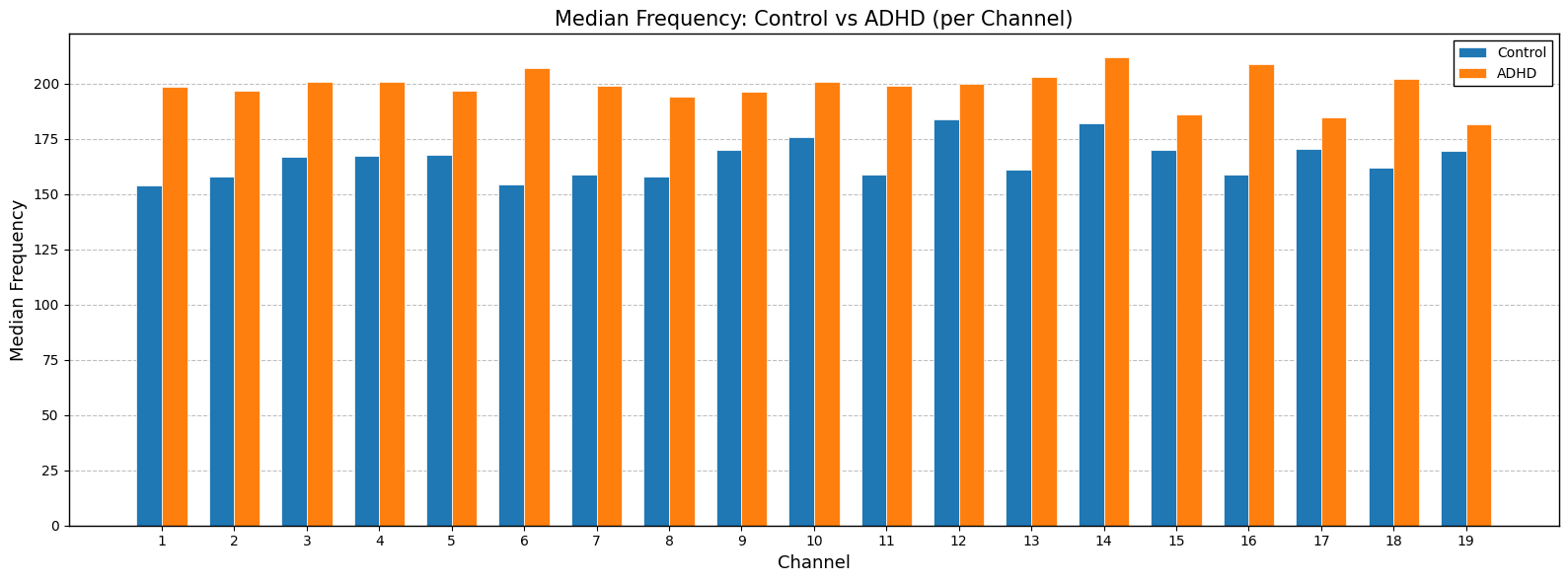}
    \caption{Median frequency of discovered CTM motifs across all EEG channels.}
    \label{fig:ctmfrequency}
\end{figure}

Across channels where motifs were detected, the frequency of motif occurrences was generally higher in the ADHD group compared to the control group (Figures~\ref{fig:opfrequency} and~\ref{fig:ctmfrequency}). This elevated motif frequency indicates increased repetitiveness and stereotypy in ADHD EEG dynamics in the subset of channels exhibiting motifs.

\begin{figure}[H]
    \centering
    \includegraphics[width=1.0\linewidth]{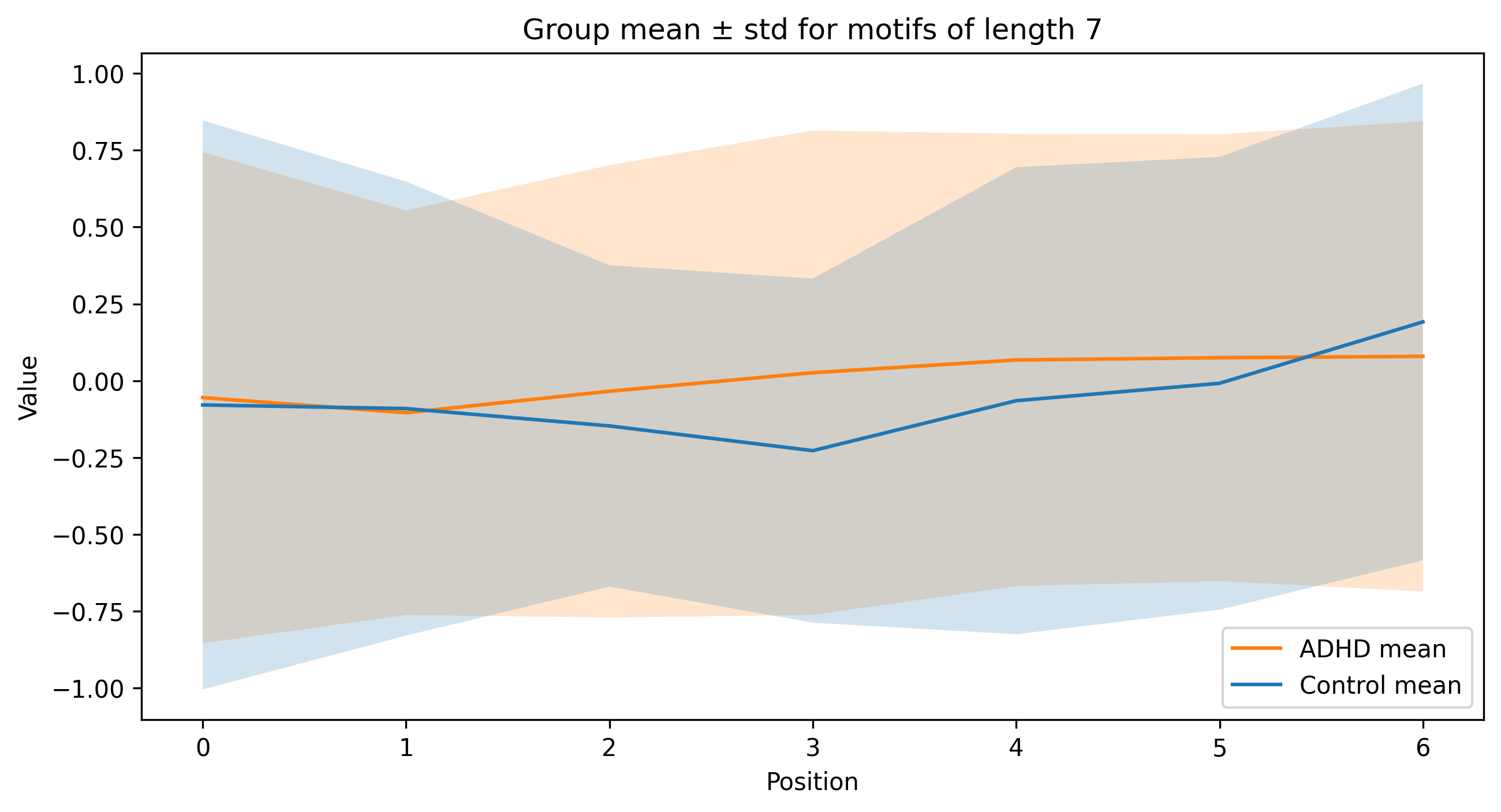}
    \caption{Detailed comparison of the mean of all discovered OPM motifs of length 7 in electrode fz (channel 17) (a.u.).}
    \label{fig:channel17len7opm}
\end{figure}

\begin{figure}[H]
    \centering
    \includegraphics[width=1\linewidth]{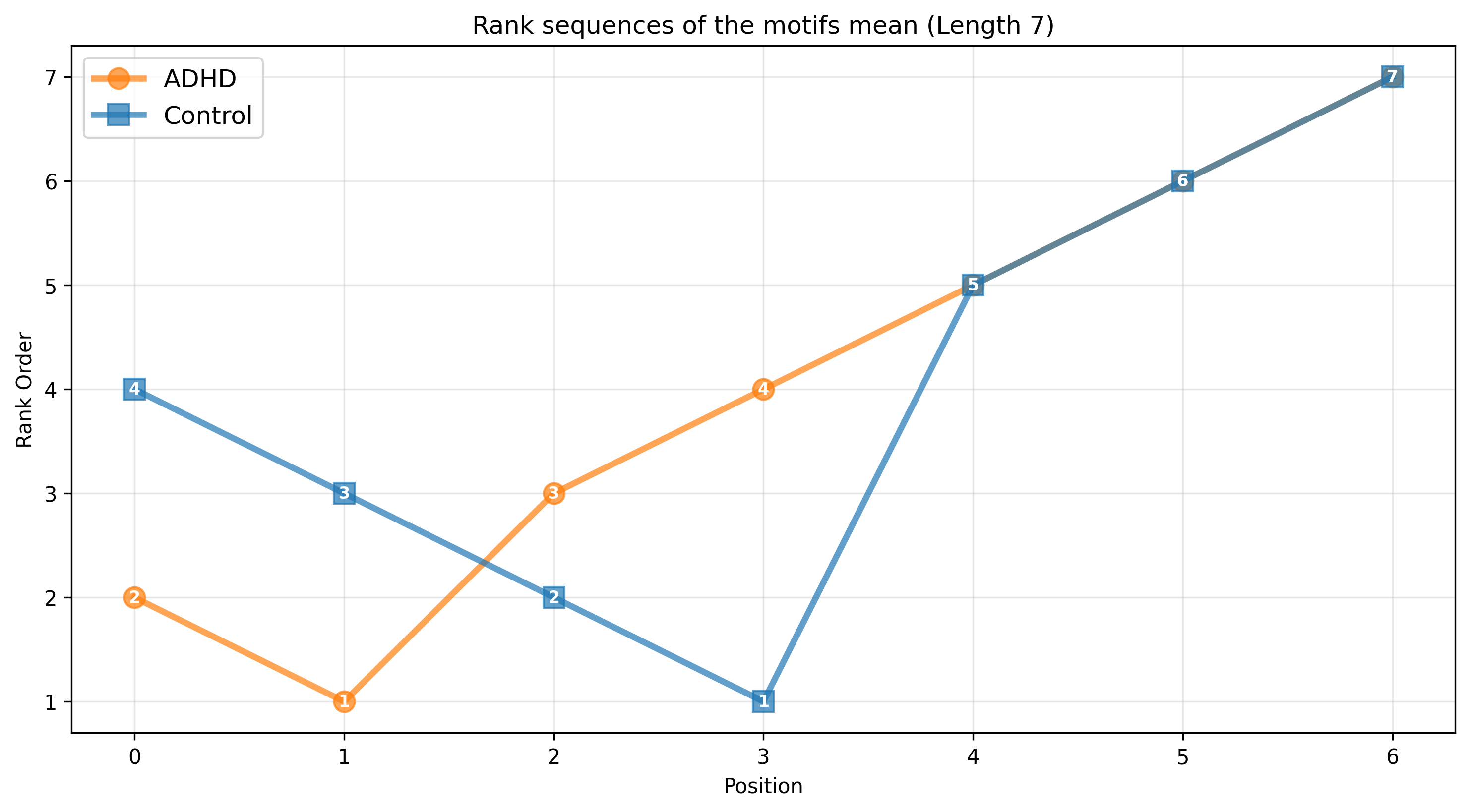}
    \caption{The ranks derived from the mean OPM motifs of length 7 in midline frontal electrode Fz (channel 17) for both groups.}
    \label{fig:ranks}
\end{figure}

Next, we show an example from a representative frontal channel (Fz, channel 17) to illustrate the magnitude of group differences in the discovered OPM motifs in Figures~\ref{fig:channel17len7opm} and~\ref{fig:ranks}.

Figure~\ref{fig:ranks} illustrates the ranks derived from the mean of OPM motifs of length 7 in midline frontal electrode Fz (channel 17) for both groups, as obtained from the preceding figure, providing an initial qualitative comparison. The sequences of length 7 mean per position of all OPM motifs for the control group and for the ADHD group are: 
\\Control: $(-0.08, -0.09, -0.15,  -0.23, -0.07,  -0.01,  0.19)$ (a.u.) with ranks $(4 ,3, 2, 1, 5, 6, 7)$.
\\ADHD: $(-0.05, -0.1, -0.03,  0.03,  0.07 , 0.075,  0.08)$ (a.u.) with ranks $(2, 1, 3, 4, 5, 6, 7)$.

\begin{figure}[H]
    \centering
    \includegraphics[width=1.0\linewidth]{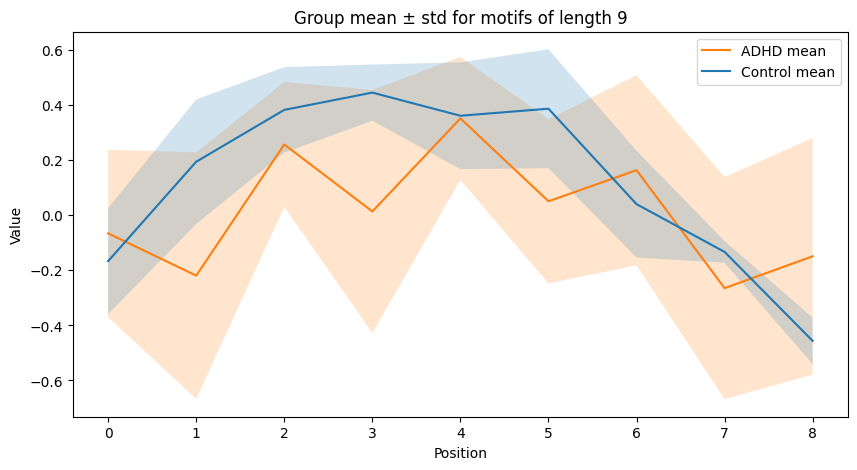}
    \caption{Detailed comparison of the mean of all discovered CTM motifs of length 9 in electrode fz (channel 17) (a.u.).}
    \label{fig:channel17len9ctm}
\end{figure}

\begin{figure}[H]
    \centering
    \includegraphics[width=1\linewidth]{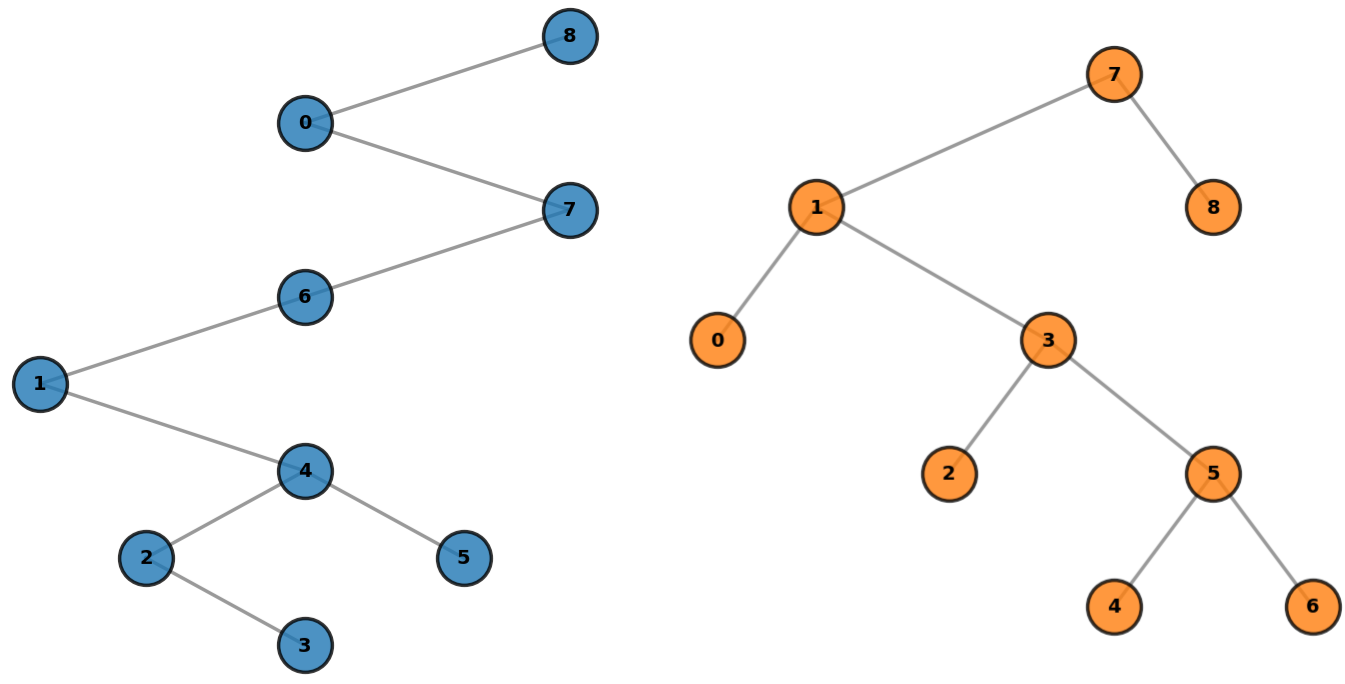}
    \caption{Left: The Cartesian tree of the mean of CTM motifs of length 9 in the control group. Right: The Cartesian tree of the mean of CTM motifs of length 9 in the ADHD group. In both trees, each node includes the position of the value.}
    \label{fig:trees}
\end{figure}

Additionally, we show an example from a representative frontal channel (Fz, channel 17) to illustrate the magnitude of group differences in the discovered CTM motifs in Figures,~\ref{fig:channel17len9ctm} and~\ref{fig:trees}.

Figure~\ref{fig:trees} illustrates the Cartesian trees derived from the mean CTM motifs of length 9 in midline frontal electrode Fz (channel 17) for both groups, as obtained from the preceding figure, providing an initial qualitative comparison. The sequences of length 9 mean per position of all CTM motifs for the control group and for the ADHD group are:
\\Control: $(-0.17,0.19,0.38,0.45,0.36,0.39,0.04,-0.13,-0.46)$ (a.u.)
\\ADHD: $(-0.07,-0.22,0.26,0.01,0.35,0.05,0.16,-0.26,-0.15)$ (a.u.)

We emphasize that, although the mean of the motifs is visualized in the figures above, the subsequent analysis does not rely on the downsampled raw EEG values themselves. Instead, OPM-based features are derived exclusively from the ordinal ranks of the motif elements, while CTM-based features depend solely on the corresponding Cartesian tree structures. Consequently, both approaches focus on relative ordering and structural shape rather than absolute signal magnitudes.

\subsection{Significant OPM Motif Features}
Order preserving matching (OPM) feature analysis revealed statistically significant group differences across multiple feature categories after FDR correction ($q<0.05$). Results are presented below, organized by feature category.

\begin{figure}[H]
\centering
\includegraphics[width=1\linewidth]{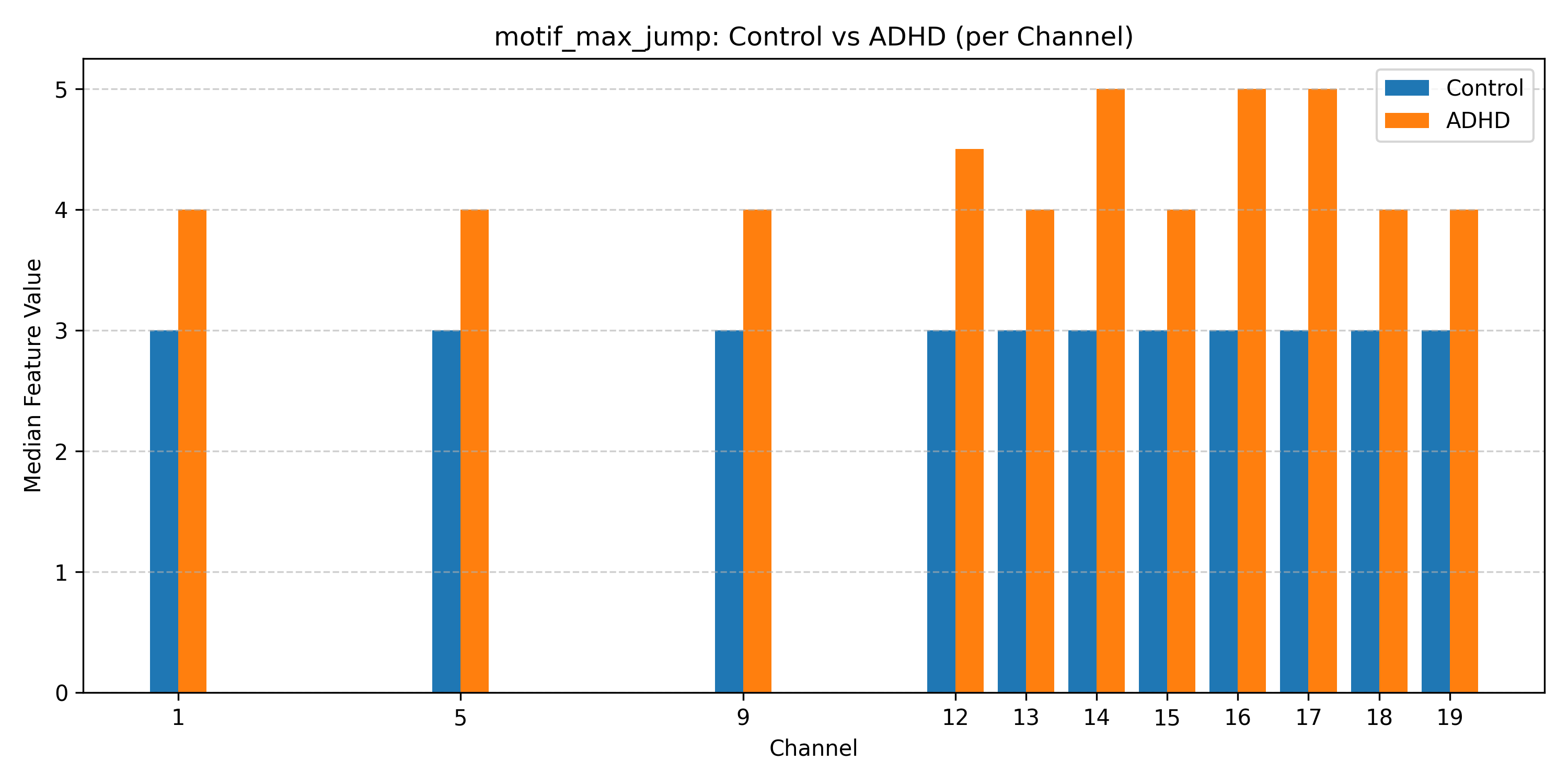}
\caption{Median max jump magnitude of all significant discovered OPM motifs. No significant group differences were observed in channels 2-4, 6-8, and 10-11 (a.u.).}
\label{fig:opmaxjump}
\end{figure}

\paragraph*{OPM Motif Gradient Features}
Analysis of gradient-based features examined motif jump-related metrics, including maximum jump, mean jump, jump variability, and normalized jump across all channels.
Motif max jump values were significantly higher in the ADHD group in all channels exhibiting statistically significant group differences, while no significant group differences were observed in channels 2-4, 6-8, and 10-11 (Figure~\ref{fig:opmaxjump}).

\begin{figure}[H]
\centering
\includegraphics[width=1.0\linewidth]{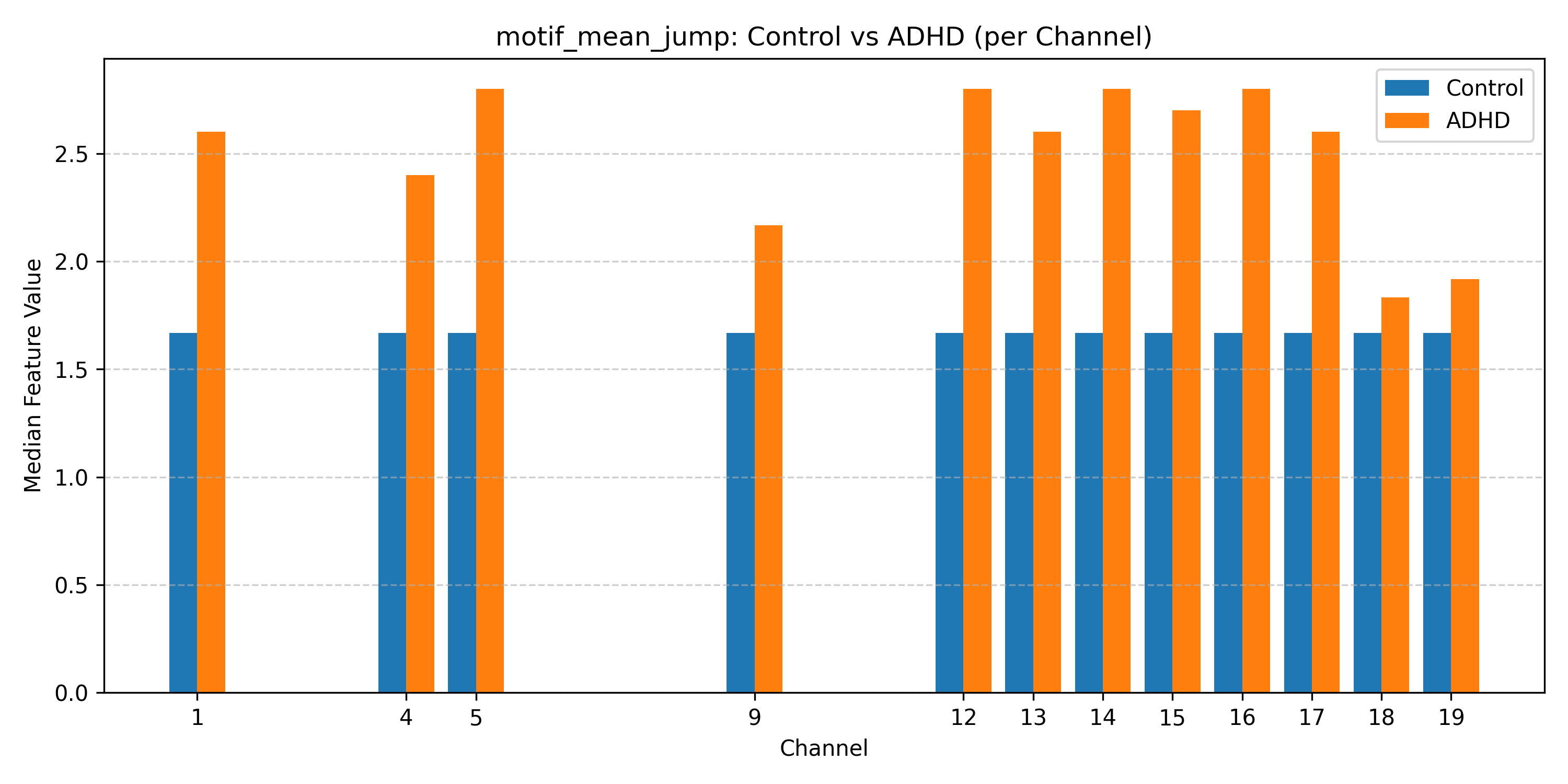}
\caption{Median mean jump of all significant discovered OPM motifs. No significant group differences were observed in channels 2-3, 6-8, and 10-11 (a.u.).}
\label{fig:opjump}
\end{figure}

Motif mean jump differed significantly between groups in the majority of channels, with higher values observed in the ADHD group; no significant group differences were detected in channels 2-3, 6-8, and 10-11 (Figure~\ref{fig:opjump}).

\begin{figure}[H]
\centering
\includegraphics[width=1.0\linewidth]{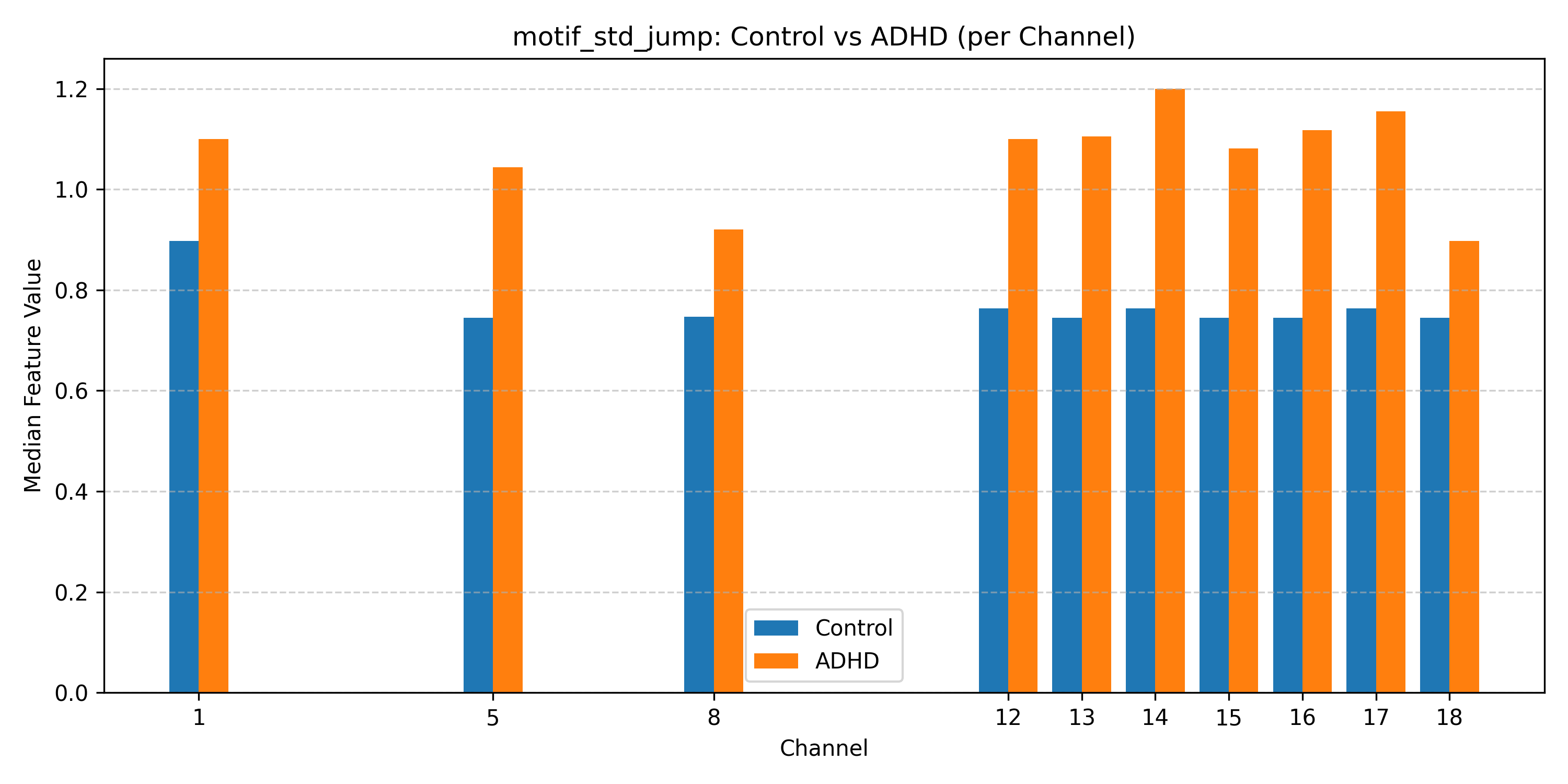}
\caption{Median standard deviation of jump magnitude for all significant discovered OPM motifs. No significant group differences were observed in channels 2-4, 6-7, 9-11, and 19 (a.u.).}
\label{fig:opstdjump}
\end{figure}

Significant group differences in motif jump variability were observed in most channels, with higher variability in the ADHD group, whereas channels 2-4, 6-7, 9-11, and 19 did not show statistically significant differences (Figure~\ref{fig:opstdjump}).

\begin{figure}[H]
\centering
\includegraphics[width=1\linewidth]{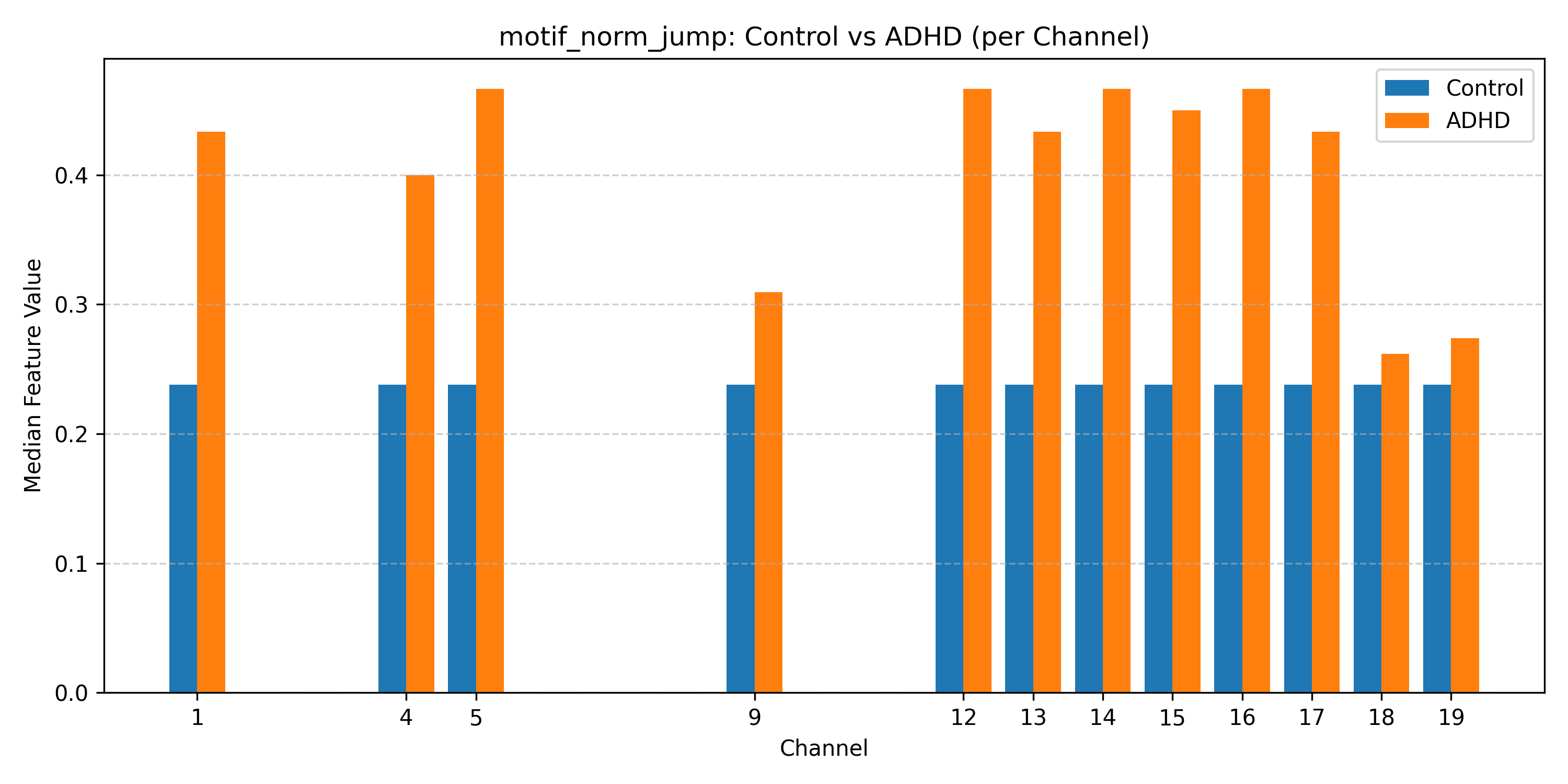}
\caption{Median normalized jump magnitude of all significant discovered OPM motifs. No significant group differences were observed in channels 2-3, 6-8, and 10-11 (a.u.).}
\label{fig:opnormjump}
\end{figure}

Normalized jump magnitude differed significantly between groups in most channels, with higher values in the ADHD group, while no significant effects were observed in channels 2-3, 6-8, and 10-11 (Figure~\ref{fig:opnormjump}).

Across all gradient-based features, ADHD participants consistently exhibited higher values in channels showing significant group differences. All four jump-related metrics (max jump, mean jump, jump variability, and normalized jump) displayed this pattern, with the majority of channels showing significant effects, though the specific non-significant channels varied slightly across features.

\begin{figure}[H]
\centering
\includegraphics[width=1\linewidth]{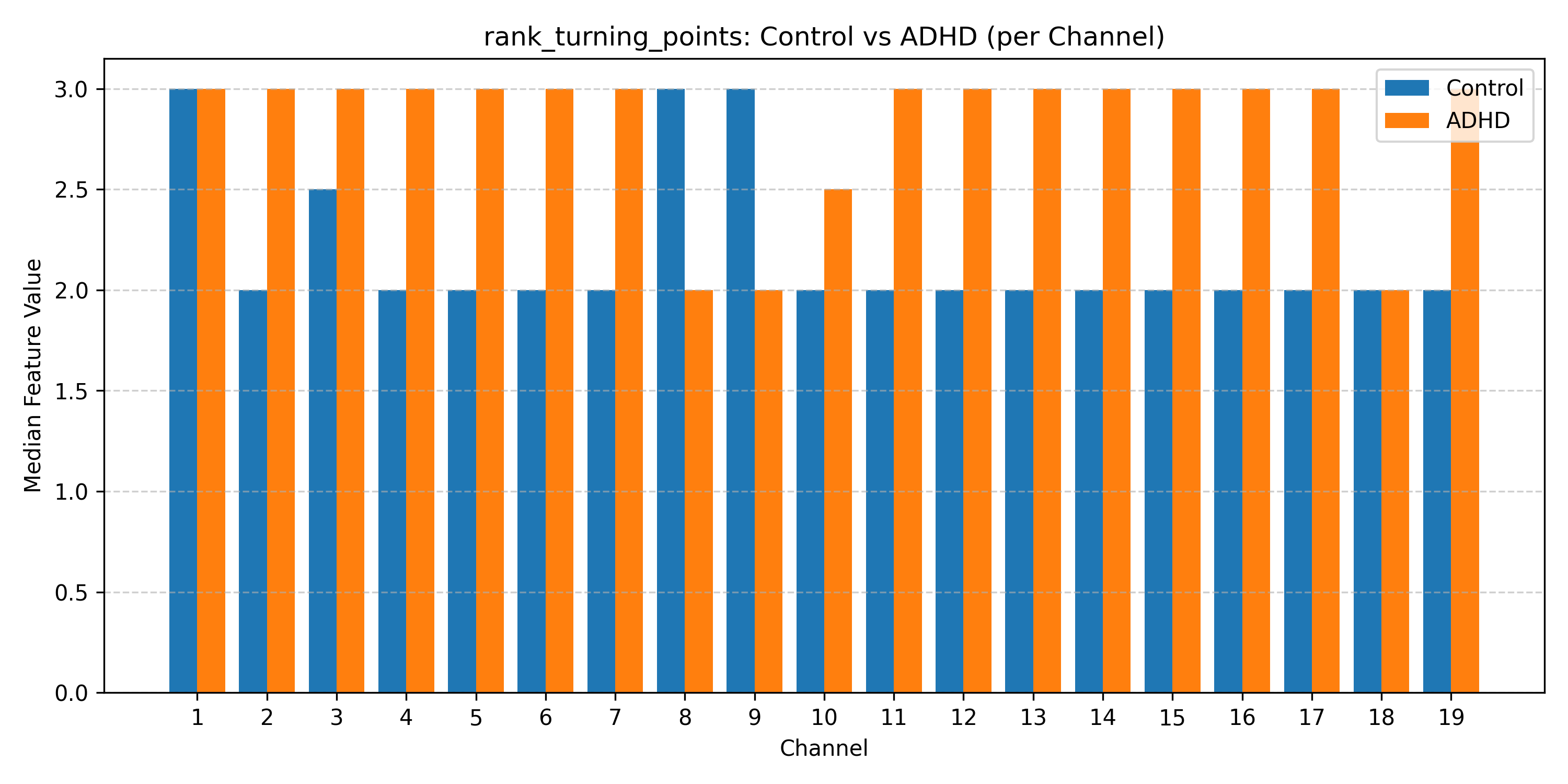}
\caption{The median value of the rank turning points count of all significant discovered OPM motifs. The rank turning points count is significant across all channels.}
\label{fig:oprank}
\end{figure}

\paragraph*{OPM Motif Rank Features}
The count of rank turning points showed significant group differences in all EEG channels, as shown in Figure~\ref{fig:oprank}. The rank turning points value is higher in ADHD participants in most channels, while channels 8 and 9 show lower values in the control group, and channels 1 and 18 show equal values in both groups.

\begin{figure}[H]
\centering
\includegraphics[width=1\linewidth]{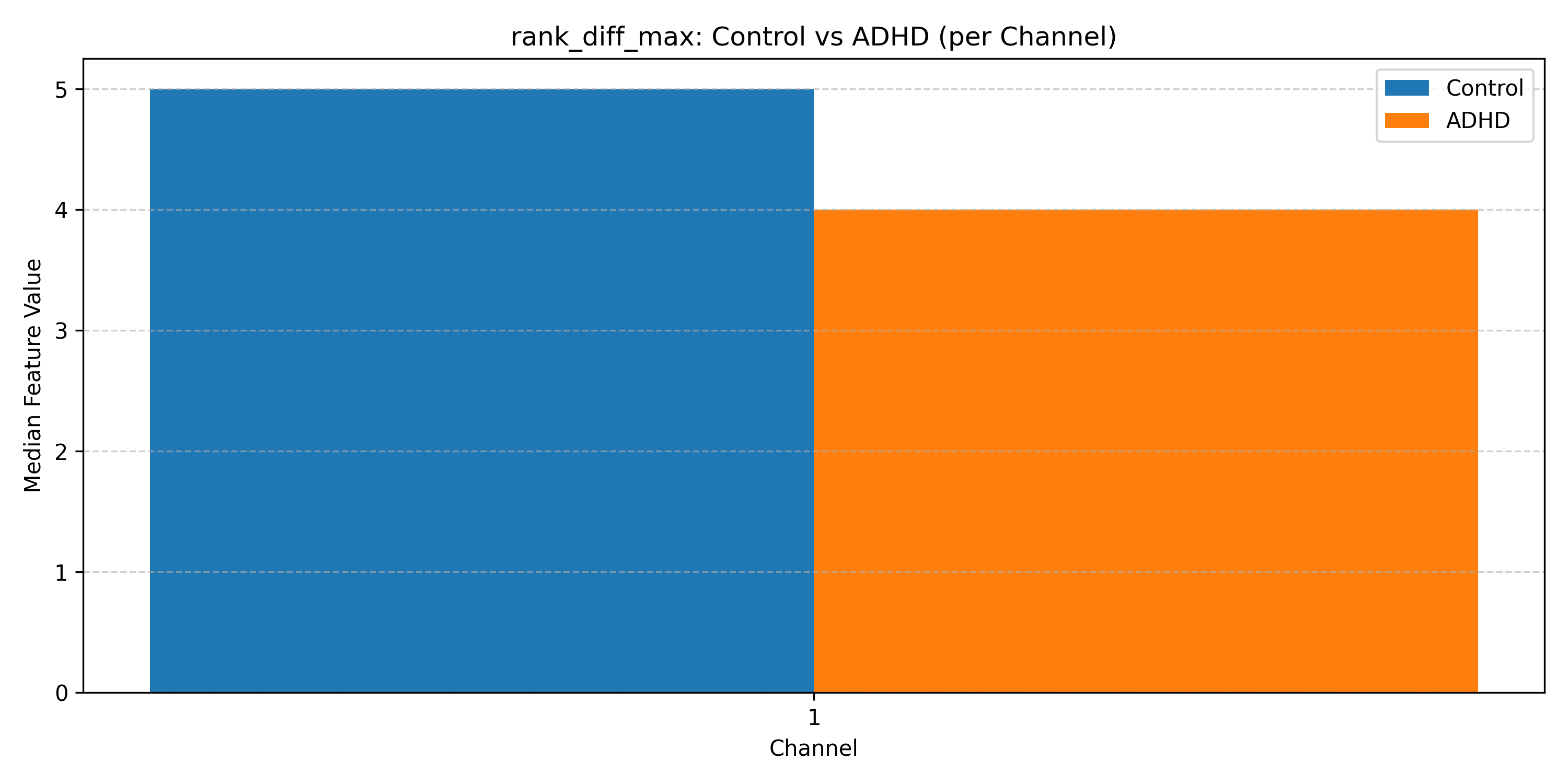}
\caption{The median value of the rank diff max of all significant discovered OPM motifs. The median value of the rank diff max is significant only in channel 1.}
\label{fig:rankdm}
\end{figure}

Rank diff max showed a significant group difference only in channel 1, with a higher value in the control group, while no other channels exhibited statistically significant effects, see Figure~\ref{fig:rankdm}.

Rank turning points exhibited significant group differences across all channels, with ADHD participants showing higher values in the majority of channels (excluding channels 1 and 18, where values were equal between groups). In contrast, rank diff max showed limited significant effects, with differences observed only in channel 1, where controls displayed higher values.

\paragraph*{Channel Locations of OPM Features}
When OPM-derived features were examined at the level of anatomically defined electrode groups (Table~\ref{tab:eeg_channels}), the following spatial distribution of significant effects was observed. In the frontal region, channels 1, 3, 4, 11, and 12 showed significant group differences in gradient-based jump measures and rank turning points, while channel 2 showed significant differences only in rank turning points, and channel 17 showed significant differences in gradient-based jump measures, rank turning points, and rank diff max. In the temporal region, both channels 13 and 14 exhibited significant differences in gradient-based jump measures and rank turning points. In the parietal region, channels 15 and 16 showed significant differences in gradient-based jump measures and rank turning points, while channels 7, 8, and 19 showed significant differences only in rank turning points. In the central region, channels 5 and 18 showed significant differences in gradient-based jump measures and rank turning points, while channel 6 showed significant differences only in rank turning points. In the occipital region, channel 9 showed significant differences only in rank turning points, while channel 10 showed no significant differences in any OPM features.

\subsection{Significant CTM Motif Features}
Analysis of CTM features revealed statistically significant group differences across both gradient and structural feature categories after FDR correction ($q<0.05$). We present these findings below, organized by feature category.

\begin{figure}[H]
\centering
\includegraphics[width=1\linewidth]{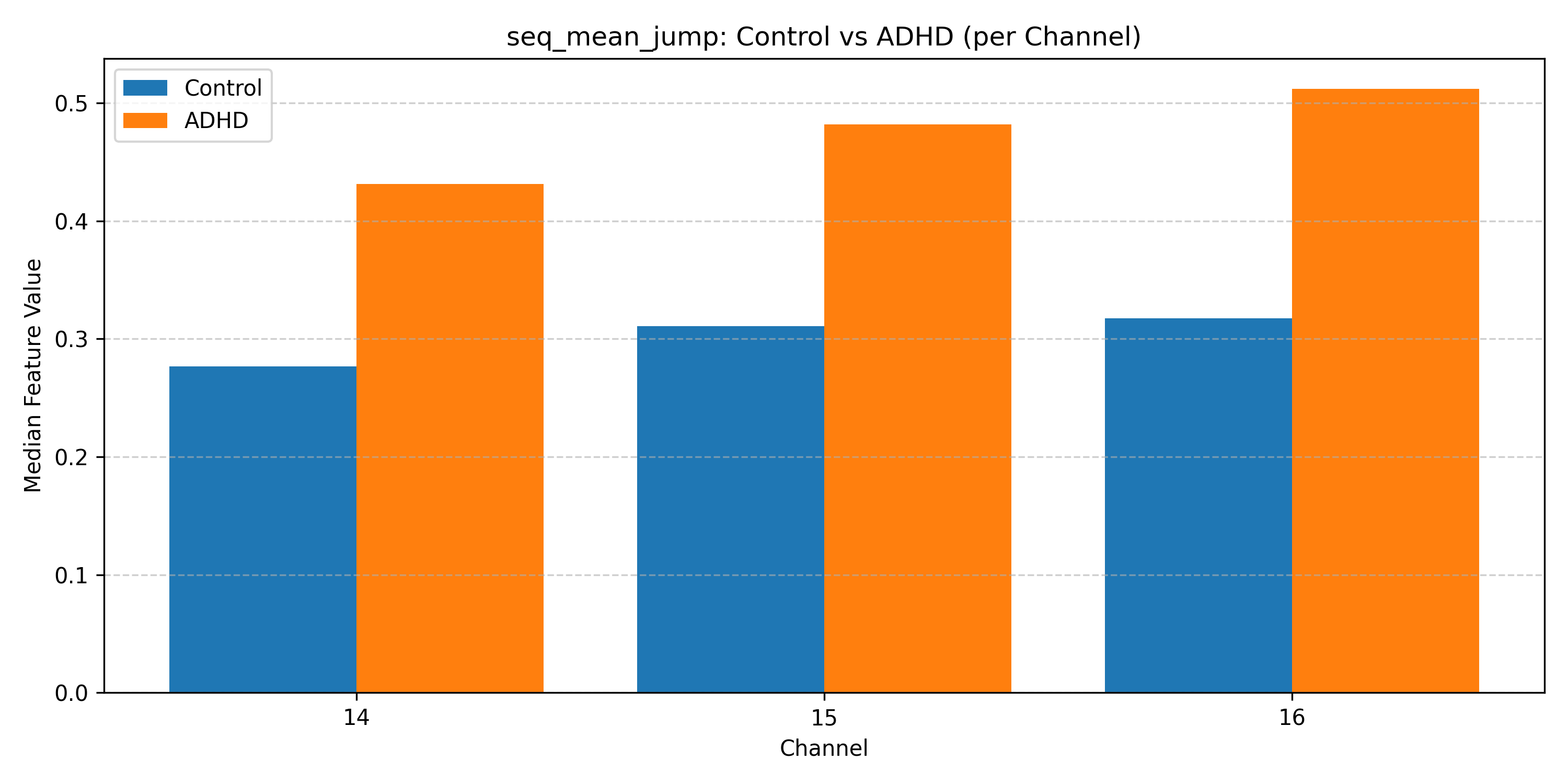}
\caption{Median mean jump in all significant CTM motifs. There are no significant motif mean jumps for channels 1-13, and 17-19 (a.u.).}
\label{fig:meanjump}
\end{figure}

\paragraph*{CTM Motif Gradient Features}
Analysis of CTM gradient features examined mean jump and max jump values across all channels. CTM motif mean jump differed significantly between groups in three channels, with no significant effects in channels 1-13 and 17-19 (Figure~\ref{fig:meanjump}).

\begin{figure}[H]
\centering
\includegraphics[width=1\linewidth]{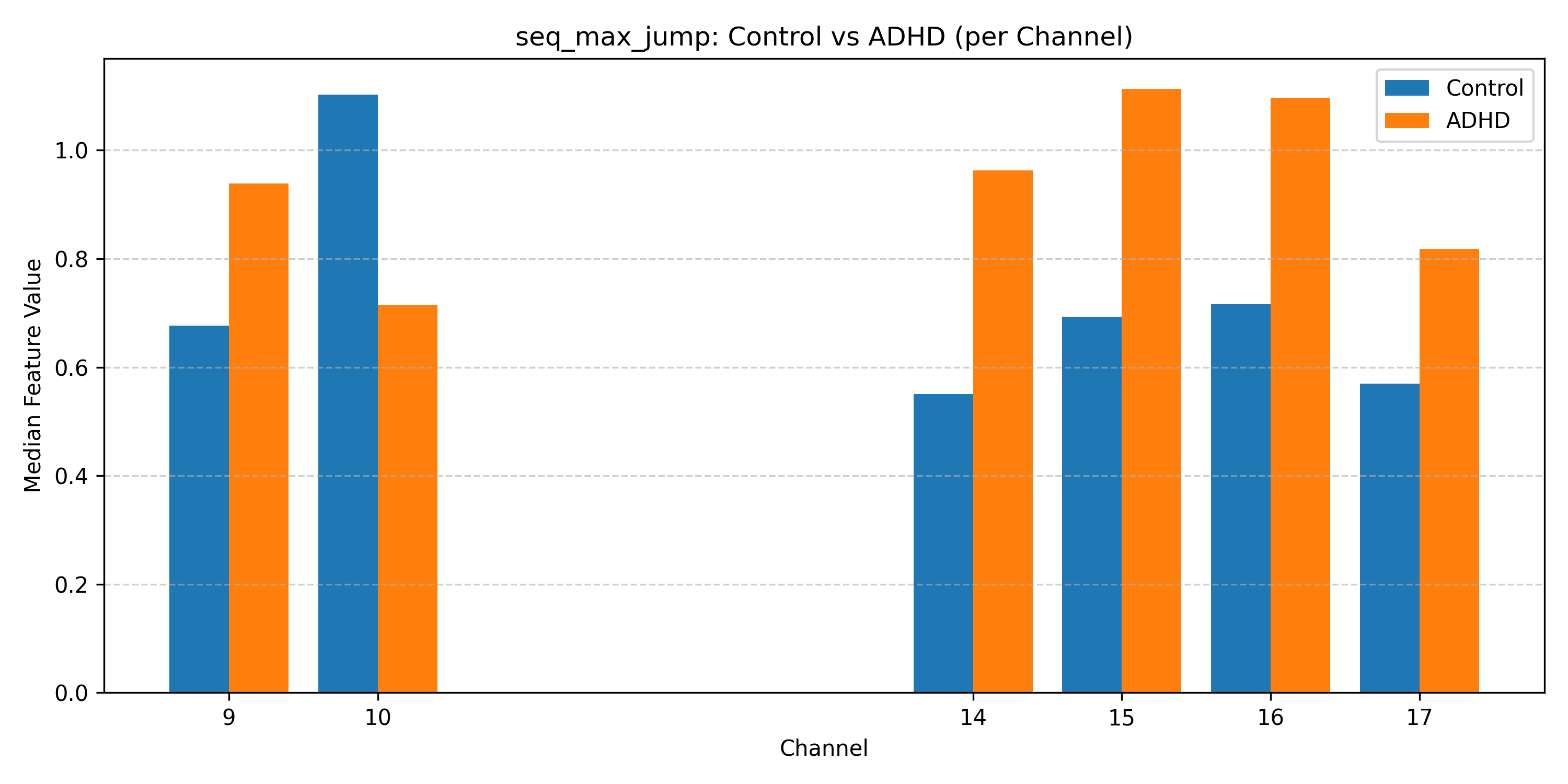}
\caption{Median max jump value in all significant CTM motifs. There are no significant motif max jumps for channels 1-8, 11-13, and 18-19 (a.u.).}
\label{fig:maxjump}
\end{figure}

Significant group differences in CTM motif max jump were observed across six channels, where in most of them the median max jump value is higher for the ADHD group, while in one channel, i.e. channel 10, the value is higher for the control group; channels 1-8, 11-13, and 18-19 did not reach significance (Figure~\ref{fig:maxjump}).

CTM gradient features showed that ADHD participants exhibited higher values in channels with significant group differences, except for channel 10, where controls displayed higher max jump values. Max jump showed more widespread significant effects (six channels) compared to the mean jump (three channels).

\begin{figure}[H]
\centering
\includegraphics[width=1\linewidth]{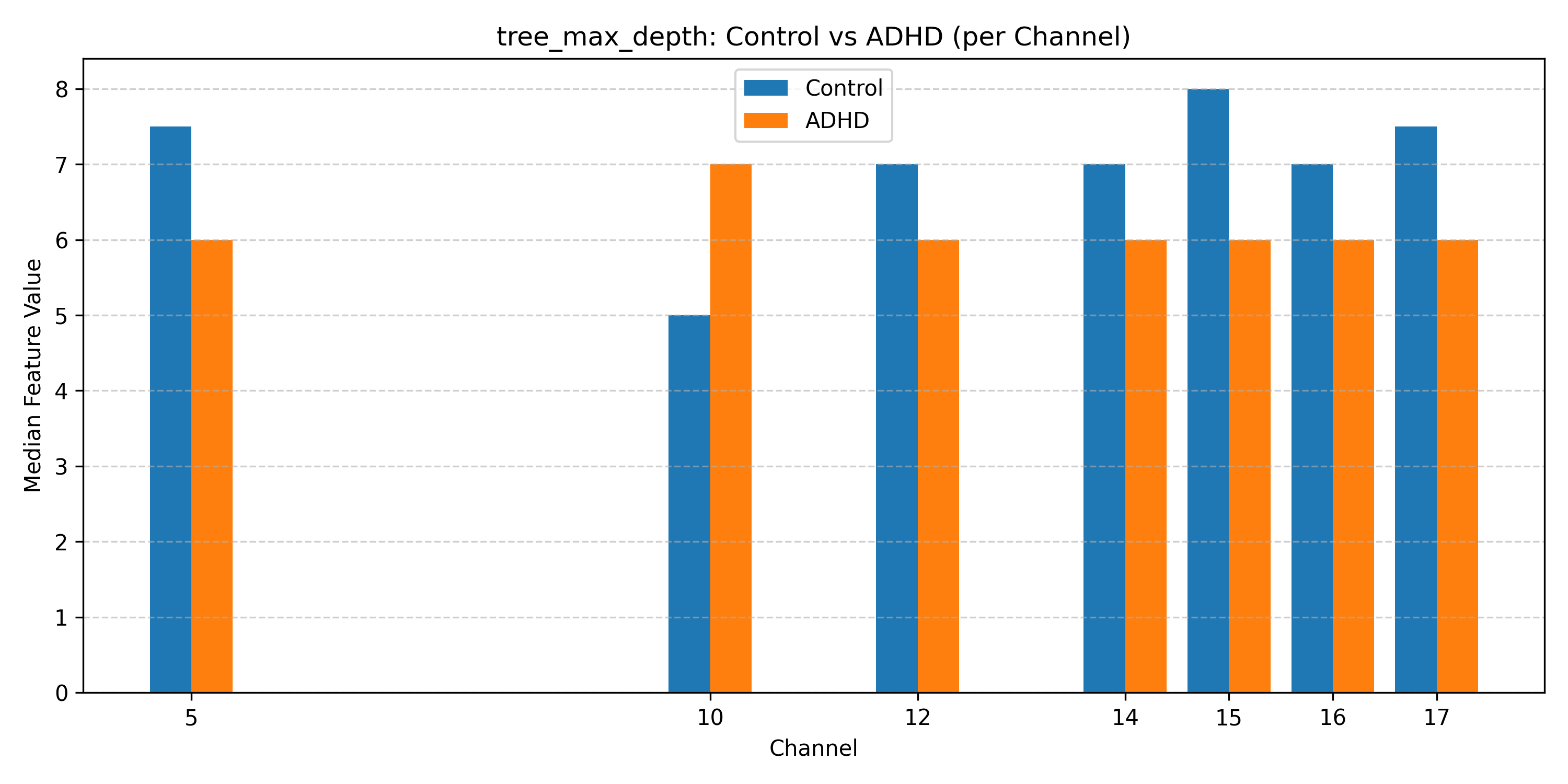}
\caption{The median of tree max depth in all significant CTM motifs. There are no significant motif tree max depths for channels 1-4, 6-9, 11, 13, and 18-19.}
\label{fig:depth}
\end{figure}

\paragraph*{CTM Motif Tree Structural Features}
Analysis of CTM tree structural features examined tree depth, leaves count, branch count, and balance measures across all channels. Note that the significant channels are similar in all tree structural features (excluding tree balance with one extra significant channel), as expected, especially for the leaves and the branches, since these two features are complementary and define all nodes in the tree.
Tree max depth showed significant group differences in seven channels, where in most of them the median tree max depth is higher for the control group, while in one channel, i.e. channel 10, the value is higher for the ADHD group; Channels 1-4, 6-9, 11, 13, and 18-19 did not exhibit statistically significant effects (Figure~\ref{fig:depth}).

\begin{figure}[H]
\centering
\includegraphics[width=1\linewidth]{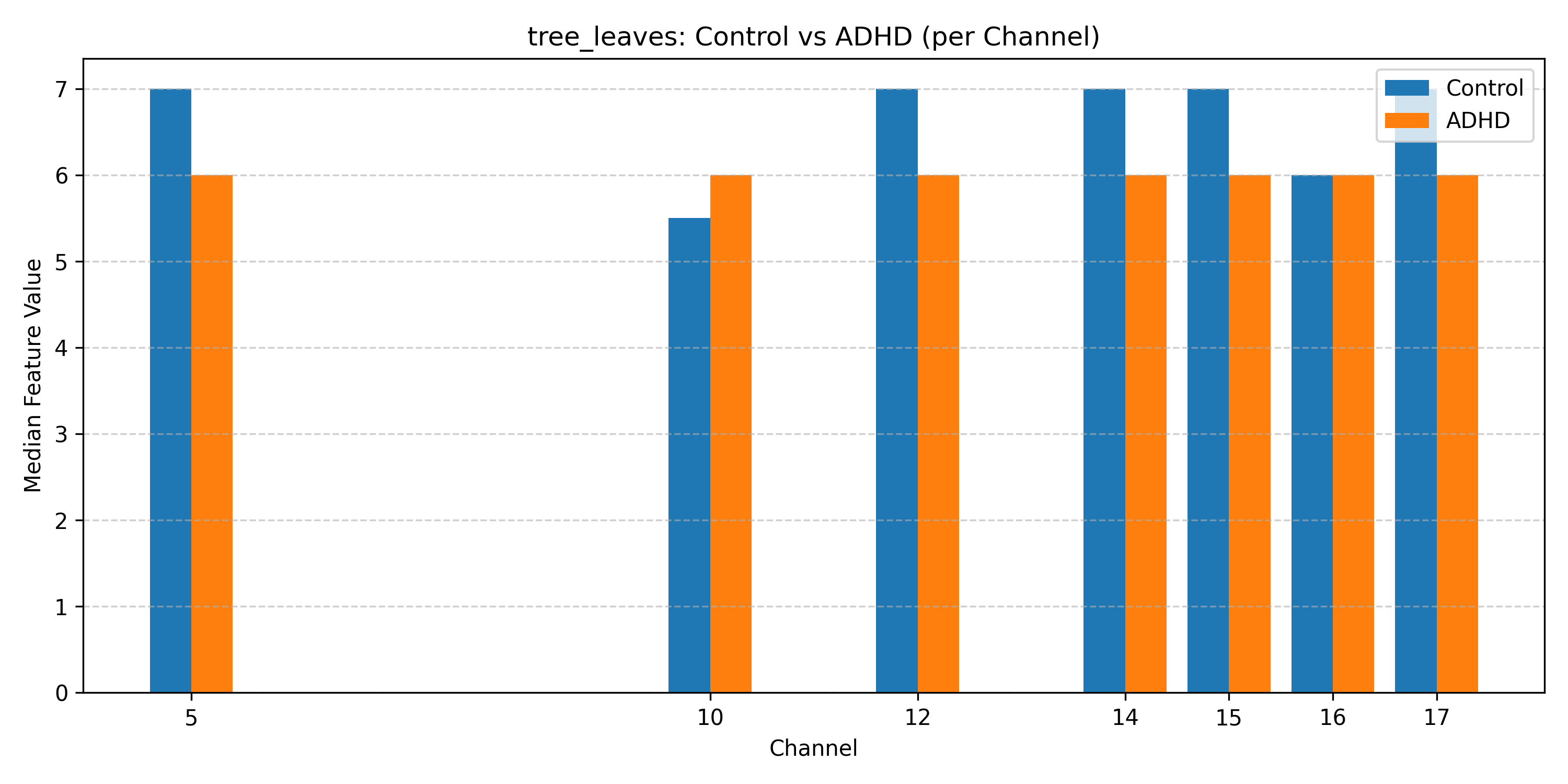}
\caption{The median of tree leaves in all significant CTM motifs. There are no significant motif tree leaves for channels 1-4, 6-9, 11, 13, and 18-19.}
\label{fig:leaves}
\end{figure}

Tree leaves count differed significantly between groups in seven of the channels, with no significant differences in channels 1-4, 6-9, 11, 13, and 18-19 (Figure~\ref{fig:leaves}); In most of them the median tree leaves count is higher (or equal in channel 16) for the control group, while only in channel 10, the value is higher for the ADHD group.

\begin{figure}[H]
\centering
\includegraphics[width=1\linewidth]{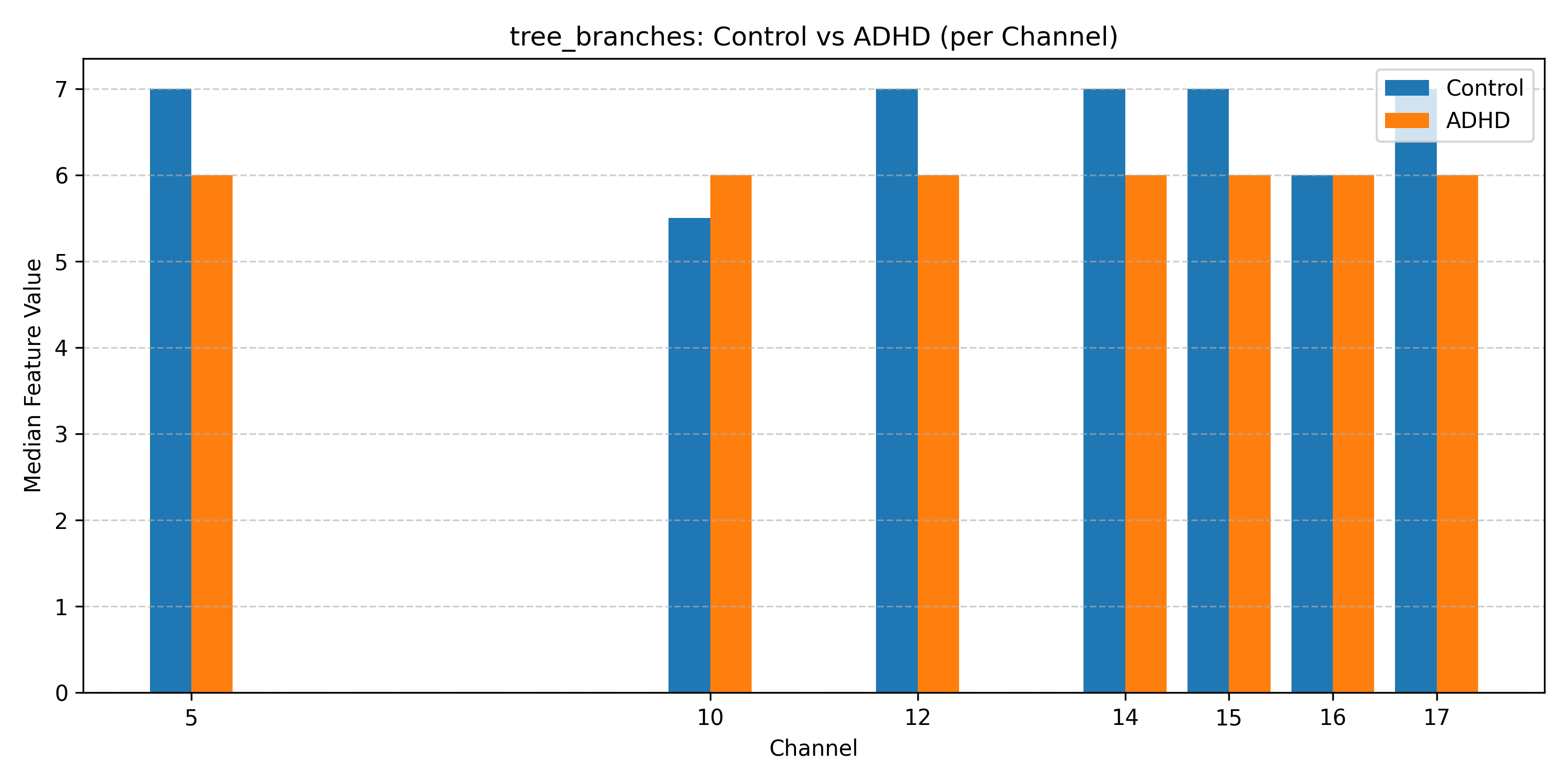}
\caption{The median of tree branches in all significant CTM motifs. There are no significant motif tree branches for channels 1-4, 6-9, 11, 13, and 18-19.}
\label{fig:branches}
\end{figure}

Significant group differences in tree branch counts were observed across seven channels, where in most of them the median tree branch count is higher (or equal in channel 16) for the control group, while only in channel 10, the value is higher for the ADHD group; channels 1-4, 6-9, 11, 13, and 18-19 did not reach statistical significance (Figure~\ref{fig:branches}).

\begin{figure}[H]
\centering
\includegraphics[width=1\linewidth]{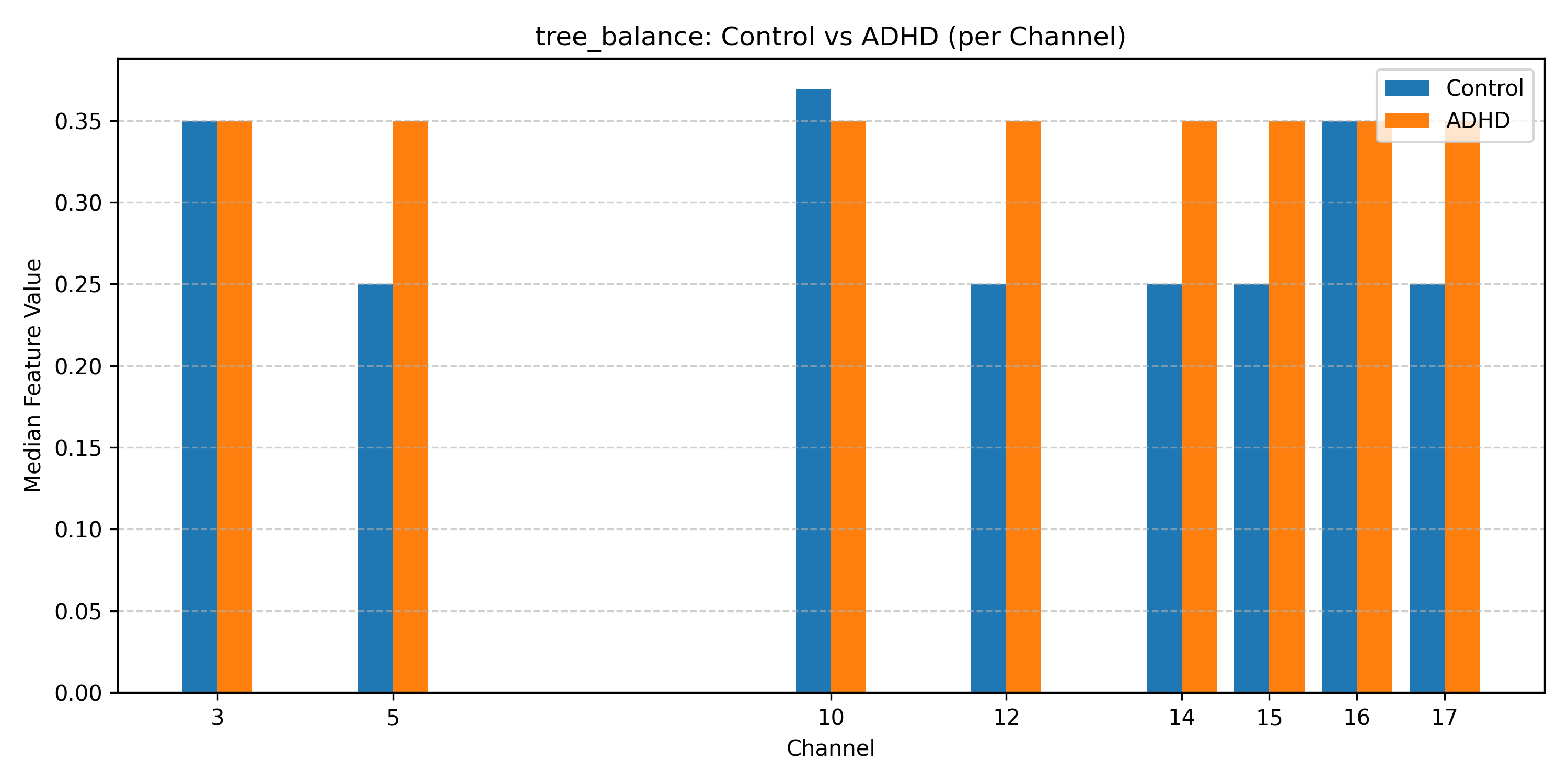}
\caption{The median of tree balance in all significant CTM motifs. There are no significant motif tree balances for channels 1-2, 4, 6-9, 11, 13, and 18-19.}
\label{fig:balance}
\end{figure}

Tree balance differed significantly between groups in eight channels, while channels 1-2, 4, 6-9, 11, 13, and 18-19 showed no significant effects (Figure~\ref{fig:balance}). In most of them, the median tree balance is higher (or equal in channels 3 and 16) for the ADHD group, while only in channel 10, the value is higher for the control group.

Structurally, in the majority of channels showing significant group differences, ADHD participants exhibited shallower trees with fewer branches and leaves compared to controls. Tree balance measures revealed a different pattern: in most significant channels, ADHD participants showed higher balance values, with channel 10 being the exception for all structural measures.

\paragraph*{Channel Locations of CTM Features}
When CTM-derived features were examined at the level of anatomically defined electrode groups (Table~\ref{tab:eeg_channels}), the following spatial distribution of significant effects was observed. In the parietal region, channels 15 and 16 showed significant group differences across all CTM features examined (mean jump, max jump, tree depth, leaves, branches, and balance), while channels 7, 8, and 19 showed no significant effects. In the temporal region, channel 14 exhibited significant differences across all CTM features, while channel 13 showed no significant effects. In the frontal region, channel 17 showed significant differences in max jump and all structural features, channel 12 showed significant differences in all structural features, channel 3 showed significant differences only in tree balance, and channels 1, 2, 4, and 11 showed no significant effects. In the central region, channel 5 showed significant differences in all structural features, while channels 6 and 18 showed no significant effects. In the occipital region, channel 10 showed significant differences in max jump and all structural features, and channel 9 showed significant differences only in max jump.
\section{Discussion}
\label{discussion}
The present study demonstrates that motif-based analysis using order preserving matching (OPM)~\cite{opm}  and Cartesian tree matching (CTM)~\cite{ctm1} provides a powerful and complementary framework for characterizing temporal dysregulation in ADHD EEG dynamics. By explicitly identifying and quantifying temporal motifs, this approach captures fine-scale waveform organization that is not accessible through traditional spectral, ERP, or global complexity measures. Together, the results reveal that ADHD-related EEG alterations are expressed as increased temporal instability at short time scales and a concurrent reduction in hierarchical organization across multiple brain regions.

\subsection{Motif Length and Frequency}
Across both OPM and CTM, motifs in the ADHD group occurred more frequently and with higher support, indicating that a smaller set of temporal patterns recurred consistently across participants. This heightened recurrence suggests increased stereotypy in neural activity. In addition, OPM motifs in the ADHD group were shorter, implying that ordinal patterns persist for briefer intervals before transitioning. Such reduced persistence may reflect instability in transient neural states associated with sustained attention, illustrating a combination of instability and repetitiveness-two features often implicated in attentional dysregulation~\cite{brennan2008neuronal,werner2020normalization}. In contrast, CTM motif lengths did not differ between groups, implying that the duration of structurally defined patterns may be preserved, even as their internal organization and hierarchical complexity are altered.
When observing the duration of the discovered motifs, i.e., the estimated time range of 0.19-0.31 seconds. Motifs of this duration correspond to theta-band activity (approximately 3-5Hz), which is associated with transitions between arousal states, including wakefulness, drowsiness, and sleep onset. Shortened motifs in this range may therefore indicate more rapid transitions between brain arousal states in ADHD, consistent with the unstable arousal regulation observed by Strauss et al.~\cite{strauss2018brain}, who found that ADHD patients showed faster declines to low EEG-vigilance stages (characterized by dominant delta- and theta-power) and reduced temporal stability of brain arousal regulation compared to controls.

\subsection{OPM-Derived Features: Fine-Scale Temporal and Ordinal Instability}
OPM provides a robust sequence-based analytical framework for identifying recurring temporal motifs in EEG data based on the relative order of signal amplitudes rather than their absolute magnitudes. This approach captures the intrinsic "shape" or structural pattern of neural fluctuations as rising or falling trends, independent of amplitude scaling or inter-individual variability. By preserving only the ordinal relationships among time-series values, OPM isolates the temporal organization of neural dynamics while minimizing sensitivity to noise and to differences in signal intensity. This property makes OPM particularly well-suited for electrophysiological data, where inter-trial variability and nonstationarity often obscure meaningful temporal structure when using conventional amplitude or frequency-based analyses.

\subsection{OPM Rank Features}
OPM Rank-based features describe the ordinal organization of motif elements, independent of amplitude scale. By focusing on the ranks of values rather than their magnitudes, these features capture the directionality and reversals of temporal trends, such as monotonic increases, decreases, or oscillatory behavior.
The most prominent effect was observed in the number of rank turning points, which was significantly elevated in the ADHD group across EEG channels, indicating a spatially widespread alteration. Increased turning point counts reflect more frequent reversals in ordinal direction, suggesting reduced temporal stability of ordinal dynamics across the scalp. Other rank-based features showed only limited or channel-specific effects, indicating that the fundamental ordinal structure is largely preserved, while its temporal modulation is more irregular in ADHD. Notably, rank diff max exhibited a localized and reversed pattern, further supporting the interpretation that ADHD-related differences are driven by frequent ordinal reversals rather than isolated extreme rank changes. This finding complements previous reports of increased neural variability in ADHD by demonstrating that instability is present even at the level of relative ordering, not only amplitude~\cite{gonen2016increased}. At the same time, the absence of widespread effects in rank correlation and rank difference means suggests that the fundamental ordinal structure remains intact. Thus, ADHD-related alterations appear to affect the dynamics of ordinal transitions rather than the global ordering itself, refining earlier interpretations of preserved but unstable neural sequencing.  

\subsection{OPM and CTM Gradient Features}
Motif gradient features quantify how rapidly the signal changes over time, providing a measure of temporal smoothness versus abruptness. Jump-based metrics capture the magnitude and variability of transitions between consecutive samples within a motif, reflecting local temporal instability.
Across both OPM and CTM, gradient features consistently indicated greater temporal instability in ADHD. Motifs in the ADHD group exhibited larger maximum jumps, higher average jump magnitudes, and greater jump variability. These findings are consistent with reports of heightened neural noise and increased trial-to-trial variability in ADHD EEG and fMRI signals~\cite{brennan2008neuronal,gonen2016increased}. Whereas traditional analyses infer instability indirectly through variability measures, gradient-based motif features provide a direct quantification of abrupt temporal changes within recurrent patterns. The convergence of OPM and CTM results in this feature category suggests that increased temporal volatility is a core characteristic of ADHD EEG dynamics, independent of whether motifs are defined by strict ordinal constraints or by hierarchical shape. 

\subsection{Spatial Distribution of OPM Effects}
When OPM-derived features were examined at the level of anatomically defined electrode groups (Table~\ref{tab:eeg_channels}), a clear qualitative spatial pattern emerged. The most pronounced ADHD-related differences were observed in frontal and temporal regions, where multiple channels exhibited significantly elevated gradient-based jump measures and increased rank turning point counts. These findings indicate greater temporal variability and more frequent ordinal reversals in motifs recorded from regions strongly implicated in executive control, attentional regulation, and cognitive flexibility~\cite{teramoto2016relation}. Parietal channels showed moderate and spatially heterogeneous effects, with some electrodes demonstrating significant group differences while others did not, suggesting regionally variable sensitivity to OPM features. In contrast, central and occipital regions exhibited weaker and less consistent effects, with several channels showing no statistically significant differences. This spatial distribution suggests that ADHD-related alterations in short-scale temporal motif organization are not globally uniform, but are more prominently expressed in frontal and temporal networks that support attentional and executive processing.

\subsection{CTM-Derived Features: Altered Hierarchical Organization of EEG Motifs}
The CTM approach extends the OPM framework by encoding the temporal structure of motifs into hierarchical tree representations that preserve essential order relationships among values while allowing for partial flexibility. 
Unlike OPM, which requires strict pairwise consistency across all elements, CTM focuses on critical order relations, such as dominant versus subordinate peaks, thereby tolerating minor deviations caused by noise, temporal jitter, or inter-individual differences. This partial order matching provides resilience against the inherent variability of EEG recordings, where small perturbations or amplitude differences can obscure structural similarity. Consequently, CTM captures the "core shape" of neural patterns with greater tolerance, facilitating the identification of consistent motifs that reflect functionally meaningful neural dynamics even under variable recording conditions.
While OPM primarily captured fine-scale temporal and ordinal instability, CTM analysis revealed pronounced group differences in the hierarchical organization of EEG motifs. Across channels exhibiting significant effects, ADHD motifs were characterized by reduced tree depth, fewer branches, and altered balance, indicating a flatter and less differentiated hierarchical structure. These findings suggest a simplification of waveform morphology, with fewer nested extrema and reduced multilevel organization.
Importantly, these structural differences were accompanied by elevated gradient-based jump measures in several channels, indicating that hierarchical simplification co-occurs with increased temporal variability within motif sequences. 

\subsection{Spatial Distribution of CTM Effects} 
Spatially, CTM-derived effects showed a non-uniform distribution across the scalp. The most consistent differences were observed in parietal and temporal regions, where multiple channels exhibited significant group differences across both structural and dynamic CTM features. Frontal and central regions showed more selective effects, primarily involving structural features and maximal temporal jumps, while occipital regions exhibited limited and feature-specific differences. This pattern suggests that ADHD-related alterations in hierarchical motif organization are regionally expressed, with stronger involvement of networks supporting attentional integration and cognitive coordination.

\subsection{Integrating OPM and CTM: Complementary Perspectives on ADHD EEG Dynamics}
Taken together, the complementary OPM and CTM analyses provide convergent evidence for increased temporal variability in ADHD EEG dynamics, while highlighting distinct but related alterations at different organizational scales. OPM primarily reveals instability in fine-scale ordinal sequencing, manifested as abrupt transitions and frequent reversals in relative ordering. CTM, in contrast, exposes a reduction in multi-level hierarchical organization, indicating diminished nesting and structural complexity of recurrent waveform patterns.
This dissociation supports a nuanced view of ADHD-related temporal dysregulation. Rather than reflecting a complete breakdown of temporal structure, ADHD EEG dynamics appear to retain basic ordinal relationships while exhibiting impaired stability and reduced hierarchical integration. Such a pattern aligns with theoretical models proposing diminished integrative processing and reduced neural flexibility in ADHD~\cite{nigg2005integrative}, and is consistent with prior evidence of reduced multiscale complexity reported using entropy, fractal, and network-based approaches~\cite{lau2022brain,catherine2022detection,fernandez2009complexity,sato2013measuring}.
By translating abstract notions of “neural noise” and reduced complexity into concrete, interpretable temporal motifs, the present findings extend existing EEG literature and demonstrate the utility of stringology-inspired methods for probing neural dynamics. Together, increased motif recurrence, elevated gradient instability, frequent ordinal reversals, and simplified hierarchical organization characterize ADHD EEG activity as structured but unstable: repetitive yet poorly integrated across time. These results highlight the potential of motif-based features as sensitive markers of temporal organization and as a complementary framework for understanding the neural mechanisms underlying attentional dysregulation.

\subsection{Limitations}
Despite the novel insights offered by this study, several limitations should be noted. First, the modest sample size and limited demographic diversity may constrain the generalizability of the observed motif-based differences; larger, more heterogeneous cohorts will be needed to examine developmental, sex-related, or subtype-specific effects. Second, participant characteristics that can influence EEG, including ADHD subtype, medication status, and comorbidities, were not systematically analyzed; future work should assess whether motif features vary across subtypes, treatment, or comorbid conditions. Third, motif discovery depends on algorithmic parameters such as motif length and similarity thresholds; although consistent patterns emerged, systematic evaluation of parameter sensitivity is needed to ensure robustness across datasets and preprocessing choices. Fourth, while OPM and CTM capture temporal order and structural regularity, they do not directly reveal underlying neural sources or functional networks; integrating motif analysis with source localization and connectivity measures will be important for linking temporal patterns to neural circuits and cognitive processes. Finally, EEG was recorded under specific conditions without manipulation of cognitive load or task demands; future studies should investigate how motif dynamics respond to behavioral states, tasks, or interventions to clarify their functional significance in ADHD.

\section{Conclusions and Future Directions}
\label{conclude}
This study demonstrates that motif-based EEG analysis using order preserving matching (OPM) and Cartesian tree matching (CTM) provides a powerful and complementary framework for characterizing temporal dysregulation in ADHD. By explicitly identifying recurrent temporal patterns and quantifying their internal structure and dynamics, this approach extends beyond traditional spectral and complexity-based analyses, which primarily summarize neural activity over extended time windows. Motif-based analysis instead captures the precise shapes and temporal persistence of short-lived waveform patterns that repeatedly emerge in the EEG signal.

Using this framework, we found that EEG dynamics in ADHD are characterized by more frequent and shorter-lived OPM motifs, indicating increased recurrence of ordinal patterns with reduced temporal persistence. This pattern reflects a temporal organization that is more repetitive and less diverse, with instability expressed through rapid transitions rather than sustained structure. Complementary CTM analysis revealed pronounced reductions in hierarchical depth, branching, and overall structural richness, pointing to diminished multiscale organization of EEG waveforms in ADHD.
By integrating motif initial, gradient, rank, and structural features, the present findings provide a unified framework that connects and extends prior EEG literature on ADHD. Increased motif recurrence corresponds to reduced temporal complexity; shorter motif persistence reflects less stable neural states; elevated gradient measures quantify neural volatility; rank turning points capture irregular ordinal dynamics; and CTM tree features indicate a tendency toward simplified hierarchical organization. Together, these results translate previously global and abstract notions of “neural noise” into concrete, interpretable temporal structures.

A clear methodological distinction emerged between the two approaches. OPM primarily revealed group differences in gradient-based features reflecting temporal instability, whereas amplitude-based statistics and most rank-based measures showed limited effects. In contrast, CTM identified significant group differences across nearly all feature categories, with motif length as the main exception. This contrast suggests that ADHD-related alterations are more pronounced in the hierarchical organization of EEG waveforms than in their ordinal sequencing: relative ordering relationships remain largely preserved, while multiscale structural complexity is consistently reduced.

Together, these findings indicate that ADHD-related EEG alterations extend beyond changes in amplitude or spectral power and instead reflect differences in the shape, stability, and internal organization of recurrent temporal patterns. By quantifying pattern recurrence, persistence, and hierarchical structure, motif-based features provide a temporally resolved perspective on neural dysregulation, linking micro-temporal EEG dynamics to broader models of network instability and reduced neural flexibility in ADHD.

Several directions for future research follow from this work. Extending motif analysis to cross-channel interactions, such as temporal lags, phase relationships, or coordinated motif occurrences across electrodes, may reveal how local temporal patterns propagate through large-scale cortical networks. Linking motif structures to specific frequency bands could clarify how canonical oscillatory mechanisms contribute to the emergence of distinct temporal patterns. Integrating motif-derived features with behavioral measures, such as reaction time variability or error rates, may further elucidate neural correlates of attentional instability and executive dysfunction. Finally, replication in larger and more diverse cohorts, across different cognitive tasks and clinical interventions, will be essential to establish the robustness, generalizability, and clinical relevance of motif-based EEG biomarkers.

Overall, OPM and CTM offer a promising methodological foundation for advancing the study of EEG temporal dynamics. By capturing temporally precise patterns that differentiate ADHD from typical development, motif-based analysis opens new avenues for understanding the neural mechanisms underlying attentional regulation. With further validation, the observed increases in motif recurrence and reductions in hierarchical complexity in ADHD may contribute to the development of objective markers for diagnosis, stratification, and treatment monitoring.

\bibliographystyle{elsarticle-num}

\end{document}